%% file: perspective.tex
\documentclass[aps,prc,showpacs,superscriptaddress,reprint,floatfix,nofootinbib]{revtex4-2}
\usepackage{xcolor}
\usepackage{url}
\usepackage{multirow}
\usepackage{makecell}
\usepackage{hyperref}
\usepackage{amsmath}

\usepackage[normalem]{ulem}

\usepackage{rotating}
\usepackage{epsfig} 
\usepackage{graphicx}
\graphicspath{ {./plots/}, {./image/}, {./figures/} }
\DeclareGraphicsExtensions{ .pdf, .png}

\usepackage{lineno}

\input{defs.tex}

\begin{document}

\title{Experimental Search for the Chiral Magnetic Effect
  in Relativistic Heavy-Ion Collisions: A Perspective}

\author{Yicheng Feng}
\email{feng216@purdue.edu}
\affiliation{Department of Physics and Astronomy, Purdue University,
  West Lafayette, IN 47907}

\author{Sergei A.~Voloshin}
\email{sergei.voloshin@wayne.edu}
\affiliation{Department of Physics and Astronomy, Wayne State
  University, Detroit, MI 48201}

\author{Fuqiang Wang}
\email{fqwang@purdue.edu}
\affiliation{Department of Physics and Astronomy, Purdue University,
  West Lafayette, IN 47907}

\date{\today}

\begin{abstract} 
The chiral magnetic effect (CME) refers to generation of the electric current along a magnetic field in a chirally imbalanced system of quarks. 
The latter is predicted by quantum chromodynamics to arise from quark interaction with non-trivial topological fluctuations of the vacuum gluonic field. 
The CME has been actively searched for in relativistic heavy-ion collisions, where such gluonic field fluctuations and a strong magnetic field are believed to be present. 
The CME-sensitive observables are unfortunately subject to a possibly large non-CME background, and firm conclusions on a CME observation have not yet been reached. 
In this perspective, we review the experimental status and progress in the CME search, from the initial measurements more than a decade ago, to the dedicated program of isobar collisions in 2018 and the release of the isobar blind analysis
result in 2022, to intriguing hints of a possible CME signal in Au+Au collisions, and discuss future prospects of a potential CME discovery in the anticipated high-statistic Au+Au collision data  at the Relativistic Heavy-Ion Collider by 2025. 
We hope such a perspective will help sharpening our focus on the fundamental physics of the CME and steer its experimental search.
\end{abstract}

\maketitle

\tableofcontents

\input{introduction}

\input{early}

\input{recent}

\section*{Acknowledgments}

This work is supported by the U.S.~Department of Energy Office of
Science, Office of Nuclear Physics under Awards No.~DE-SC0012910 (YF, FW) and DE-FG02-92ER40713 (SV).

\bibliography{./ref}


\end{document}

%% file: defs.tex
\usepackage{xspace}
\usepackage[normalem]{ulem}

\newcommand{\snn}{\sqrt{s_{_{\textsc{nn}}}}}

\newcommand{\avfd}{{\textsc{avfd}}}
\newcommand{\ampt}{{\textsc{ampt}}}
\newcommand{\epos}{{\textsc{epos}}}
\newcommand{\hydjet}{{\textsc{hydjet}}}
\newcommand{\hijing}{{\textsc{hijing}}}

\newcommand{\dndeta}{\ensuremath{\mathrm{d}N_\mathrm{ch}/\mathrm{d}\eta}}
\newcommand{\pt}{p_{T}}
\newcommand{\gevc}{GeV/$c$}
\newcommand{\Nch}{N_{\rm ch}}

\newcommand{\poi}{{\textsc{poi}}}
\newcommand{\obs}{{\text{obs}}}
\newcommand{\cme}{{\textsc{cme}}}

\newcommand{\dg}{\Delta\gamma}
\newcommand{\dgv}{(\Delta\gamma/v_2)}
\newcommand{\dgamma}{\Delta\gamma}
\newcommand{\sms }{\scriptstyle}
\newcommand{\ru}{\rm Ru+Ru}
\newcommand{\zr}{\rm Zr+Zr}
\newcommand{\Ru}{\rm Ru+Ru}
\newcommand{\Zr}{\rm Zr+Zr}

\newcommand{\dd}{\Delta\delta}
\newcommand{\fcme}{f_{\textsc{cme}}}
\newcommand{\minv}{m_{\rm inv}}

\newcommand{\deta}{\Delta\eta}
\newcommand{\dphi}{\Delta\phi}
\newcommand{\tphi}{\tilde{\phi}}
\newcommand{\srp}{{\textsc{rp}}}
\newcommand{\ssp}{{\textsc{sp}}}
\newcommand{\spp}{{\textsc{pp}}}
\newcommand{\sep}{{\textsc{ep}}}
\newcommand{\sos}{{\textsc{os}}}
\newcommand{\sss}{{\textsc{ss}}}

\newcommand{\gos}{{\ensuremath{\gamma_\sos}\xspace}}
\newcommand{\gss}{{\ensuremath{\gamma_\sss}\xspace}}
\newcommand{\gab}{{\gamma_{\alpha\beta}}}

\newcommand{\two}{{\{2\}}}
\newcommand{\EP}{\{\sep\}}

\newcommand{\zdc}{\{{\textsc{zdc}}\}}
\newcommand{\tpc}{\{{\textsc{tpc}}\}}

\newcommand{\vtwopp}{v_2^\spp}
\newcommand{\vtwosp}{v_2^\ssp}
\newcommand{\dgpp}{\dg^\spp}
\newcommand{\dgsp}{\dg^\ssp}
\newcommand{\dgcme}{\dg_{\textsc{cme}}}

\newcommand {\mean}[1]   {\langle{#1}\rangle}

\newcommand {\comment}[1]   {\iffalse {#1} \fi}

\newcommand{\psiep}{\ensuremath{\psi_{\textsc{EP}}\xspace}}
\newcommand{\psirp}{\ensuremath{\psi_{\textsc{RP}}\xspace}}

\newcommand{\psisp}{\ensuremath{\psi_{\textsc{SP}}\xspace}}

%% file: introduction.tex
\section{The Chiral Magnetic Effect}
\label{sec:intro:cme}

The universe came into being with a Big Bang. 
After breaking of the electroweak symmetry by the Higgs mechanism~\cite{Englert:1964et,Higgs:1964pj}, quarks and leptons  acquired mass.
Although not massless, the quarks were light and the chiral (left- and right-handed) symmetry was approximately preserved~\cite{Nambu:1960xd}.  
The universe was then, for a period of approximately 10~$\mu$sec, in a state of freely roaming quarks and gluons, the so-called quark-gluon plasma (QGP)~\cite{Shuryak:1978ij}.  
The universe expanded, cooled down, and underwent the hadronization phase transition with spontaneous chiral symmetry breaking; the quarks became ``dressed'' by gluons and grouped into massive hadrons.

The universe following evolution until the present day is relatively well understood, but the details of its evolution in the QGP phase and the exact mechanisms responsible for the chiral symmetry breaking are poorly known. 
In Quantum Chromodynamics (QCD), a non-abelean quantum gauge theory describing the interaction of quarks and gluons, the vacuum possesses a non-trivial structure with potential barriers separating vacua with different topological properties. 
The transitions between different topological sectors, described by instantons and (at non-zero temperature) sphalerons, and accompanied by a flip in quark chirality, play a  key role in the chiral symmetry breaking and supposedly in mass generation~{\cite{tHooft:1986ooh}} (for a review, see~\cite{Shuryak:1980tp,Schafer:1996wv}).

The chirality flips of the quarks during the topological transitions create a chirality imbalance, a difference in the numbers of left- and right-handed fermions. The chirality imbalance is predicted, in the presence of a strong magnetic field, to induce an electric current along the magnetic field.
This phenomenon is called the chiral magnetic effect (CME)~\cite{Kharzeev:2007jp,Fukushima:2008xe}.  
The induced electric current can be described as ~\cite{Yin:2015fca,Jiang:2016wve,Shi:2017cpu}
\begin{equation}\label{eq:j5}
    \vec{\mathbf{J}} = \frac{Q_f^2}{2\pi^2}\mu_5\vec{\mathbf{B}}\,,
\end{equation}
where $Q_f$ is the quark electric charge, and
$\mu_5$ is an axial chemical potential that quantifies the amount of chirality imbalance.  

\begin{figure}
  \includegraphics[width=0.85\linewidth]{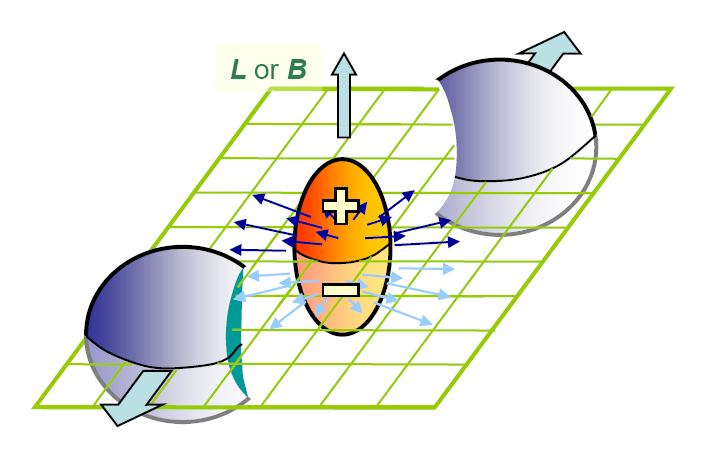}
  \caption{ (Color online) Schematic view of a heavy-ion collision. 
    The charge separation happens along the magnetic field indicated by the upward arrow.
    The direction of the charge separation fluctuates in accord with the sign of the topological charge.
    Drawing is from Ref.~\cite{STAR:2009tro}.}
  \label{fig:cartoon}
\end{figure}

Relativistic heavy-ion collisions provide an ideal environment to search for the CME.  
In high-temperature QGP created in those collisions, the chiral symmetry is restored and the transitions between topologically different vacuum states may be enhanced~\cite{Kharzeev:1999cz,Kharzeev:2004ey,Kharzeev:2007jp}.  
In addition, a large magnetic field on the order of $B\sim m_\pi^2/e \sim 10^{18}$~Gauss can be created (here $e$ is the elementary charge and $m_\pi$ is the pion mass)~\cite{Skokov:2009qp,Voronyuk:2011jd,Deng:2012pc}.  
The cartoon in Fig.~\ref{fig:cartoon} depicts a noncentral heavy-ion collision, where the reaction plane (RP), spanned by the impact parameter direction and the beam direction, is indicated by the grid parallelogram. 
The magnetic field, created mostly by spectator protons of the colliding nuclei, points perpendicular to the RP.

The magnetic field in heavy-ion collisions is expected to quickly fade off as the two spectator remnants recede from each other. 
However, the decay time can be significantly prolonged depending on the electric conductivity of the QGP~\cite{Tuchin:2013apa,McLerran:2013hla,Li:2018ufq,Huang:2022qdn}.
With plausible parameters for the axial current density and the time evolution of the magnetic field, anomalous (viscous) hydrodynamic calculations suggest that an appreciable CME can manifest itself in relativistic heavy-ion collisions~\cite{Yin:2015fca,Shi:2017cpu,Jiang:2016wve}.

Experiments at the Relativistic Heavy-Ion Collider (RHIC) at Brookhaven National Laboratory in the U.S.~and at the Large Hadron Collider (LHC) at CERN in Europe~\cite{Adcox:2004mh,Adams:2005dq,Arsene:2004fa,Back:2004je,  Muller:2012zq,Roland:2014jsa} are set to study the properties of the QGP.
It is found that the QGP is strongly coupled and behaves like a nearly perfect fluid~\cite{Gyulassy:2004zy,Shuryak:2008eq}.  
Experimental searches for the CME in relativistic heavy-ion collisions at RHIC and the LHC have persisted since the mid-2000s. 
Significant progress has been made, while many challenges related to the treatment of background effects remain open.  
In this Perspective, we review these developments and challenges and discuss prospects of a potential CME discovery in high-statistics heavy-ion data to be available in the next years.  
For further reading, the reader is referred to extensive reviews on the subject in Refs.~\cite{Kharzeev:2013ffa,Kharzeev:2015znc,Huang:2015oca,Hattori:2016emy,Zhao:2018ixy,Zhao:2019hta,Li:2020dwr,Kharzeev:2020jxw,Kharzeev:2024zzm}.  

While we focus on the CME in this perspective, there exist several other chiral effects, such as the chiral magnetic wave (CMW) and the chiral vortical effect (CVE), which shares similar observable techniques as the CME's. For discussion of these closely related phenomena, see Refs.~\cite{Kharzeev:2010gd,Burnier:2011bf,Landsteiner:2011iq,Burnier:2012ae,Yee:2013cya,Kharzeev:2015znc,STAR:2015wza,ALICE:2015cjr,CMS:2017pah}.

\section{Experimental Observables and Methods}
\label{sec:methods}

\subsection{Gamma correlator}
\label{sec:methods:gamma}

A main signature of the CME is the charge separation along the magnetic field and  perpendicular to the RP.
In a noncentral heavy-ion collision, the overlap interaction region is of an almond shape (see Fig.~\ref{fig:cartoon}).  
The pressure buildup inside the fireball drives a rapid expansion of the system, strongest in the direction of the highest pressure gradient (short axis)~\cite{Ollitrault:1992bk}.  
This results in an anisotropic azimuthal ($\phi$) distribution of the final-state particles, which is most often characterized with the help of Fourier expansion~\cite{Voloshin:1994mz},
\begin{eqnarray}
  \label{eq:Fourier}
  \frac{dN_\pm}{d\phi} \propto 1 &+& 2v_1\cos(\phi-\psirp)
  + 2v_2\cos2(\phi-\psirp) + \cdots
  \nonumber\\
    &+& 2a_{1\pm}\sin(\phi_\pm-\psirp) + \cdots \,.
\end{eqnarray}
In this expression, we have also added sine terms to account for a possible particle distribution asymmetry in the direction of the magnetic field, on average perpendicular to the reaction plane.
The subscripts $\pm$ denote the electric charge sign of the particle.  
The coefficients $v_n$ quantify the strengths of the corresponding anisotropic flow: the parameter $v_1$ is often called directed flow, and $v_2$ elliptic flow. 
Neglecting the effects of the electromagnetic interactions, the $v_n$ coefficients are charge independent.  
The parity ($\mathcal{P}$)-odd first harmonic sine term in Eq.~(\ref{eq:Fourier}) with coefficients $a_{1+} = -a_{1-}$ is used to describe the main effect of charge asymmetry caused by the CME.
Due to the random nature of the topological charge fluctuations, the averages of $\mean{a_\pm}$ over many events vanish.
Only the correlation measures $\mean{a_{1+}a_{1-}}=-a_1^2$ and $\mean{a_{1\pm}a_{1\pm}}=a_1^2$ can be finite.

\begin{figure}
  \includegraphics[width=0.85\linewidth]{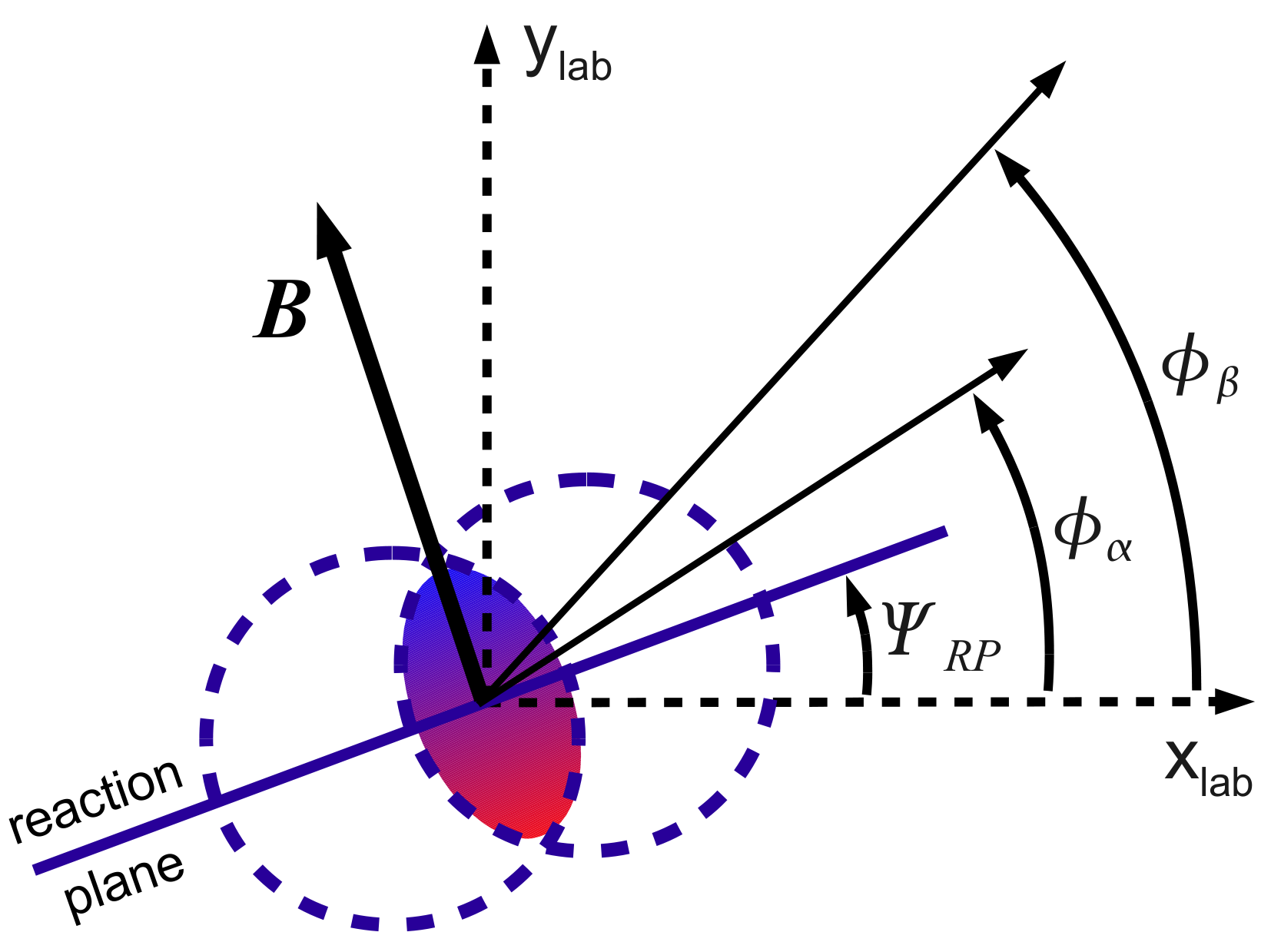}
  \caption{(Color online) Schematic view of the transverse plane in a collision of two heavy nuclei -- one emerging from and one going into the page. 
  The azimuthal angles of the reaction plane and produced
  particles with charges $\alpha$ and $\beta$ as used in
  Eqs.~(\ref{eq:gamma}) and (\ref{eq:c3}) are depicted here. 
  Drawing is from Ref.~\cite{STAR:2009wot}.}
  \label{fig:planes}
\end{figure}

A variable capturing the CME physics that has been most commonly used in experimental CME searches is the charge-dependent three-point azimuthal correlator~\cite{Voloshin:2004vk},
\begin{eqnarray}
  \label{eq:gamma}
  \gamma_{\alpha\beta} &\equiv& \langle \cos(\phi_\alpha +\phi_\beta
  -2\psirp) \rangle
  \\  
  &=& \langle \cos\tphi_\alpha\, \cos\tphi_\beta\rangle
  - \langle \sin\tphi_\alpha\,\sin\tphi_\beta  \rangle
  \nonumber
  \\ &=& [\langle v_{1,\alpha}v_{1,\beta} \rangle + B_{\rm IN}] 
  - [\langle a_{\alpha} a_{\beta} \rangle + B_{\rm OUT}]\,.
  \nonumber
\end{eqnarray}
Here, $\phi_\alpha$ and $\phi_\beta$ are the azimuthal angles of two particles of interest (POI),  $\psirp$ is the RP angle (see Fig.~\ref{fig:planes}), and $\tphi=\phi-\psirp$. 
The averaging is done over all particles in the event and over all events.
The subscripts $\alpha$ and $\beta$ indicate the charge signs of the POIs.  
$B_{\rm IN,OUT}$ represent contributions from $\mathcal{P}$-even background processes.  
As seen from the expression Eq.~(\ref{eq:gamma}) the gamma correlator represents the difference in correlations of the particle emission direction projected onto the RP and on the normal to the RP -- the direction of the magnetic field.  
Thus, the correlator eliminates all background contributions not related to the RP orientation. 
The remaining background, $B_{\rm IN}-B_{\rm OUT}$, is suppressed approximately by a factor of $v_2$.

Due to initial density fluctuations, the particle azimuthal distribution in each event might not be symmetric about the RP. 
In that case the anisotropic flow part in expansion Eq.~(\ref{eq:Fourier}) is substituted by a similar one with $\psirp$ changed to $\psi_n$, the $n$th-order harmonic flow plane. 
Those planes are also called participant planes (PP), as they are determined by the initial matter distribution geometry.  
The magnetic field direction, determined mostly by spectator protons, should be much less affected by fluctuations in the participant zone, and more strongly correlated to the real RP.
The flow coefficients measured relative to different planes, wherever important, are denoted with superscript, e.g.~$v_n^\textsc{rp}=\mean{\cos n(\phi-\psirp)}$ and
$v_n^\textsc{pp}=\mean{\cos n(\phi-\psi_n)}$. 
In this perspective, we also discuss azimuthal correlations relative to the plane determined by the spectator nucleons, called the spectator plane (SP), in which case the notation would be $v_n^\textsc{sp}=\mean{\cos n(\phi-\psisp)}$. 
Similar notation will be used for the $\gamma$ correlator.

If  CME were the only correlations among particles, then the gamma correlator values measured with the opposite charge-sign (OS) particle pairs $\gos$, and that for the same sign (SS) pairs $\gss$,  would be opposite in sign ($\gos>0$ and $\gss<0$) and same in magnitude, but this relationship can be modified by contributions from correlations of non-CME origins.  
The charge-independent background contribution to the correlator can be eliminated by taking the difference between the $\gos$ and $\gss$ correlators~\cite{Voloshin:2004vk},
\begin{equation}\label{eq:dg}
    \dg \equiv \gos - \gss\,.
\end{equation}
The remaining charge-dependent non-CME contributions represent the main/major problem in the CME search. 
They are discussed in detail in the following sections.

It has become a common approach to quantify the CME signal in terms of the CME-fraction, usually presented in percent, which is the relative contribution of the CME correlation to the measured gamma correlator value,
\begin{equation}
  \fcme\equiv \frac{\dgcme}{\dg}\,.
\end{equation}
As we show below, the most precise techniques allow one to estimate such a fraction with an accuracy of a few percent.

Several other observables for the CME search have also been proposed~\cite{Ajitanand:2010rc,Magdy:2017yje,Feng:2018chm,Feng:2020cgf,Bzdak:2011np,Sun:2018onk,Tang:2019pbl,Feng:2019pxu,Li:2024pue,Li:2024pue}.  
Most of them are similar to the $\gamma$ correlators. 
Some of the observables were investigated in Ref.~\cite{Choudhury:2021jwd} and found to have similar or worse sensitivities to the CME signal than the $\dg$ observable.
The $\dg$ correlator and its several variants, as discussed in Sect.~\ref{sec:new}, appear to be the most promising and economic observable to search for the CME. 
We will thus concentrate on $\dg$ in this perspective.

\subsection{Major backgrounds}
\label{sec:intro:backgrounds}

\subsubsection{Reaction-plane dependent (flow induced)  background}
\label{sec:RPdependent}

While the gamma correlators by construction scale down the contribution from non-CME correlations by approximately a factor of $v_2$, the difference between $B_{\rm IN}$ and $B_{\rm OUT}$ can still be significant.
This contribution is often discussed in terms of  ``flowing clusters''~\cite{Voloshin:2004vk,Wang:2009kd,Bzdak:2009fc,Schlichting:2010qia} and can be presented as 
\begin{equation}
\label{eq:cluster}
  \frac{B_{\rm IN}-B_{\rm OUT}}{B_{\rm IN}+B_{\rm OUT}}= 
v_{2,{\rm cl}}\frac{\mean{\cos(\phi_\alpha+\phi_\beta-2\phi_{\rm cl})}}
{\mean{\cos(\phi_\alpha-\phi_\beta)}},
\end{equation}
where $v_{2,{\rm cl}}$ is the elliptic flow, $\phi_{\rm cl}$ is the emission azimuthal angle, and the POIs are two decay products of the cluster.
This background is caused by two-particle correlations due to cluster decays in presence of elliptic flow, and often referred to as ``flow-induced'' background.

Resonances and jets are natural candidates for the clusters responsible for background. 
Model calculations (see Sect.~\ref{sec:first}) show that only those could not explain the observed signal. 
Contribution from another type of clusters, often referred to as local charge conservation (LCC), originates from an interplay of locally produced charges, e.g.~by gluon conversion to quark-antiquark pair, with strong elliptic flow. 
Corresponding Blast-Wave model calculations with reasonable parameters show that the LCC effect could be of the order of the observed signal in the experiment.

\subsubsection{Nonflow and  reaction-plane independent background}
\label{sec:RPindependent}

The harmonic planes $\psi_n$ cannot be measured in experiment, and are usually estimated from  azimuthal distributions of the final-state particles~\cite{Poskanzer:1998yz},
\begin{equation}\label{eq:psi}
  \psi_{n,\sep}
  = \frac{1}{n} \arctan\left(
  \frac{\mean{\sin n \phi}}{\mean{\cos n \phi}} \right)\,,
\end{equation}
where the average is taken over all particles in an event in a given momentum region.  
The experimentally reconstructed $\psi_{n,\sep}$, called the event plane (EP), deviates from the geometry defined $\psi_n$ due to the finite number of particles used in its reconstruction, the effect of which can be quantified by the so-called EP resolution, $R_\sep \equiv\mean{\cos  n(\psi_{n,\sep}-\psi_n)}$. 
The latter can be evaluated from the correlations between different EPs, e.g. by the sub-event method~\cite{Poskanzer:1998yz}.
The $v_n$ coefficients, corrected for the EP resolution, are calculated as
\begin{equation}
    v_n\EP = \mean{\cos n(\phi-\psi_{n,\sep})}/R_\sep\,.
\end{equation}
Similarly, the measurements of $v_n$ can be done with  a two-particle correlation method as 
\begin{equation}\label{eq:v22}
    v_n\two = \sqrt{\mean{\cos n(\phi_1-\phi_2)}}\,,
\end{equation}
where both particles are taken from the same momentum region.  
In both methods, it is assumed that the main contribution to the correlations between the two particles or a particle and the EP is due to anisotropic flow.  
While these methods could provide good estimates for flow coefficients, the results might be biased by a contribution from correlations not related to the RP orientation, such as due to inter- and intra-jet correlations and resonance decays, often referred to as ``nonflow''.
In this Perspective, we use the well-established notation of $v_n\{...\}$ with curly brackets to denote the experimental measurements obtained with specific method, e.g.~$v_2 \two$ would mean elliptic flow measured via 2-particle correlations. 
Having in mind that both particles are correlated with the participant plane, the full notation in this case could be $v_2^\spp \two$, but for simplicity, if not confusing, the superscript will be omitted. 
On the other hand, $v_n$ without curly brackets is referred to the true collective flow, as those in Eq.~(\ref{eq:Fourier}), where the harmonic planes are defined by the initial-state geometry.

Similar to Eq.~(\ref{eq:v22}), the $\gamma$ correlator can also be calculated by correlating the particles $\alpha$ and $\beta$ to a third particle $c$~\cite{Voloshin:2004vk,STAR:2009tro,STAR:2009wot},
\begin{equation}\label{eq:c3}
    C_{3,\alpha\beta}=\mean{\cos(\phi_\alpha+\phi_\beta-2\phi_{c})}\,,
\end{equation}
so that
\begin{equation}\label{eq:c3v2}
   \gab = C_{3,\alpha\beta} / v_{2,c}\,.
\end{equation}
The azimuthal direction of particle $c$ in this case plays the role of $\psi_2$ with the resolution equal to the elliptic flow parameter $v_{2,c}$ of particle $c$.

The three-particle correlator used to calculate gamma via Eq.~(\ref{eq:c3}) contains  correlations not related to the RP, e.g.~from back-to-back dijets, with particles $\alpha$ and $\beta$ originating from one jet and the particle $c$ from the other jet. 
It is known that the  intra-jet particle correlations are strongly charge-dependent; therefore, such RP-independent background has to be taken care of by other means. 
This background contribution, representing three-particle cumulants, should scale as an inverse of the multiplicity squared, and thus has a strong centrality dependence.
As we will see in Sect.~\ref{sec:first} this type of background is dominant in  peripheral nuclear collisions and makes the measurements in small systems, Sect.~\ref{sec:small}, not practical for the CME search.

We also note that $v_{2,c}$ measurement might be biased by nonflow, which in turn affects the $\gamma$ correlator calculated by Eq.~(\ref{eq:c3v2}).
The effect of nonflow correlations is discussed in more detail in Sect.~\ref{sec:new:refine}.

\subsubsection{Vector meson spin alignment}
\label{sec:alignment}

A possibility of spin polarization along the global angular momentum direction in heavy-ion collisions was first suggested in Refs.~\cite{Liang:2004ph,Voloshin:2004ha}.  
$\Lambda$-hyperon global polarization has been measured by STAR~\cite{STAR:2007ccu,STAR:2017ckg} to be of the order of a fraction of a percent at the highest RHIC energy and up to a few percent at lower collision energies.  
The global hyperon polarization is well described by hydrodynamic models accounting for vorticity in the system. 
More recently, a nonzero spin alignment of the $\phi$ meson relative to the second-order harmonic plane has also been  reported~\cite{STAR:2022fan}.  
The magnitude of the reported spin alignment ($\sim1\%$) appears to be orders of magnitude larger than expected from any processes related to vorticity and thus is unlikely related to the observed hyperon polarization~\cite{Yang:2017sdk,Sheng:2019kmk,Xia:2020tyd,Gao:2021rom,Muller:2021hpe}.  
One explanation for the observed spin alignment discussed in Refs.~\cite{Sheng:2019kmk,Sheng:2022wsy,Sheng:2022ffb,Sheng:2023urn}, 
is the large fluctuations of the correlated strangeness ($s$) and anti-strangeness ($\bar{s}$) color fields, though that mechanism would not explain the measured $\pt$ and centrality dependence of the effect.  
It has been argued~\cite{Shen:2022gtl} that $\rho^0$-meson spin alignment, if similar to that reported for the $\phi$-meson, can contribute to charge separation $\dg$ measurements due to the anisotropic distribution of the daughter $\pi^+$ and $\pi^-$ in the $\rho^0$ rest frame relative to the RP.  
This effect, if exists, is unlikely to be related to the magnetic field, and thus could be treated similarly to other RP-dependent backgrounds discussed in Sect.~\ref{sec:RPdependent}.

\subsection{Experimental methods and techniques}
\label{sec:methods:techniques}

Over the years, many different experimental approaches have been proposed to separate the signal from the background. 
Those could be subdivided into two groups: estimates of the background from the measurements where no signal is expected, and extraction of the signal by the measurements where the relative contributions of the signal and background could be varied in controlled manner by modification of the background or the signal. 
As we discuss below, different approaches have different systematic uncertainties, and not all of them are up to the precision needed at present time. 
In the rest of this section we briefly discuss the strong and weak sides of different techniques, but leave a quantitative comparison to later sections presenting the existing experimental results.

\subsubsection{Higher and mixed harmonics}
\label{sec:methods:mixed}

In this approach the estimate of the background effects is based on the observation that while the gamma correlator as defined in Eq.~(\ref{eq:gamma}) includes the contribution from the CME and background, similarly defined correlators involving higher harmonics, e.g.~doubled harmonic correlator $\mean{\cos2(\phi_\alpha+\phi_\beta-2\psi_4)}$~\cite{Voloshin:2011mx}, includes only background contribution, though corresponding not to the second but to the fourth harmonic flow. 
The preliminary results obtained by this method were reported in~\cite{Voloshin:2012fv}. 
Unfortunately, it is rather difficult to precisely calculate the relative difference in the backgrounds corresponding to different harmonic correlators, and thus to obtain any reliable estimates of the CME effect.

A more radical proposal was to relate the background in the gamma correlator to the values of the two-particle $\delta$ correlator: 
\begin{eqnarray}
    \label{eq:delta}
    \delta_{\alpha\beta} &\equiv&
    \mean{\cos(\tphi_\alpha-\tphi_\beta)} \nonumber
    \\ &=& 
    \mean{\cos\tphi_\alpha\cos\tphi_\beta}
    + \mean{\sin\tphi_\alpha\sin\tphi_\beta} \,.
\end{eqnarray}
The background correlations contributing almost equally to $\mean{\cos\tphi_\alpha\cos\tphi_\beta}$ and $\mean{\sin\tphi_\alpha\sin\tphi_\beta}$, largely cancel in $\dg$ but add up in $\dd\equiv\delta_\sos-\delta_\sss$ where they overwhelm any CME signal.
The authors of Ref.~\cite{Bzdak:2012ia} attempted to relate the background in
$\dg$ to $\dd$ through an approximate relationship  
\begin{eqnarray}
  \label{eq:gamma-delta}
  \gamma_{\alpha\beta} &\equiv& \mean{\cos(\tphi_\alpha
    +\tphi_\beta)} 
  \nonumber\\
  &\approx&\mean{\cos(\tphi_\alpha
    -\tphi_\beta)\cos2\tphi_\beta}
  \nonumber\\
  &\approx& v_2\mean{\cos(\tphi_\alpha-\tphi_\beta)}
  =v_2\delta_{\alpha\beta} \,,
\end{eqnarray}
and arrived at the following equations~\cite{Bzdak:2012ia}:
\begin{subequations}\label{eq:HF}
    \begin{align}
        \dg &\approx  v_2 F - H \,, \\
        \dd &= F + H \,, 
    \end{align}
\end{subequations}
where $F$ is the ``flow'' background, $H$ is the CME signal, and $F\gg H$.  
Then Eqs.~\ref{eq:HF} could be used to extract the CME signal. 
However, one could note that the factorization in Eq.~(\ref{eq:gamma-delta}) is mathematically invalid because the angles $\tphi_\alpha-\tphi_\beta$ and $2\tphi_\beta$ are not independent.
The correct factorization of the terms is done in Eq.~(\ref{eq:cluster}).
For the corresponding experimental measurements and quantitative discussion of this approach, see Sect.~\ref{sec:first:kappa}.

In the same spirit, one can construct another correlator 
$\gamma_{123}\equiv\mean{\cos(\phi_\alpha+2\phi_\beta-3\psi_3)}$~\cite{  CMS:2017pah,ALICE:2020siw}.  
Since it is defined with respect to the third-order harmonic plane $\psi_3$, there is no CME contribution to the $\gamma_{123}$ correlator.  
One could then take the $\dg_{123}$ correlator as a background estimator for $\dg_{112}\equiv\dg$ and perform the factorization~\cite{CMS:2017pah,ALICE:2020siw} $\dg_{123} \approx
v_3 \dd$ to obtain a qualitative estimate of the CME.   
Again, such a factorization is mathematically unjustified, and this approach could not be used for a quantitative estimate of the CME signal or background.

\subsubsection{U+U collisions}
\label{sec:methods:UU}

The (ultra)central uranium collision could provide a unique opportunity to measure the flow-induced background in the CME search~\cite{Voloshin:2010ut}. 
Due to the nonspherical shape of the uranium nuclei, the elliptic flow in such collisions could vary by more than a factor of two and on average is significantly larger than that in Au+Au collisions. 
Elliptic flow is the smallest in the so-called tip-tip collisions and the largest in body-body collisions. 
The magnetic field, which is mostly due to spectator protons, is small in most central collisions and rather similar in U+U and Au+Au collisions.
The three-particle RP-independent background is also small in such a collision because of the large multiplicity. 
Thus, the dominant contribution to the gamma correlator comes from elliptic flow induced correlations as given by Eq.~\ref{eq:cluster}.
Then selecting the events corresponding to different orientations of the uranium nuclei, e.g.~by applying event-shape engineering technique~\cite{Schukraft:2012ah}, one can vary the elliptic flow and in this way estimate the background contribution to the gamma correlator as a function of the elliptic flow.

\subsubsection{Isobar collisions}
\label{sec:methods:isobar}

In order to separate the CME contribution to $\dg$ from the background, one can also try to vary the signal contribution, keeping the background relatively constant.  
To vary the signal, it was proposed by one of us~\cite{Voloshin:2010ut} to use the isobar collisions. 
The isobar nuclei have the same number of nucleons and different numbers of protons. 
Thus, the magnetic field (and, correspondingly, the CME signal) is expected to be different, whereas the background due to flow is expected to be very similar.  
Early model calculations~\cite{Deng:2016knn} suggest that about $4\times10^9$ minimum bias (MB) events for each isobar species would be sufficient to reach a $5\sigma$ significance if the CME signal is about 10\% of the inclusive $\dg$ measurement, a number consistent with measurements in Au+Au collisions.
Event-by-event \avfd\ calculations also conclude that a CME observation in isobar collisions is likely given the plausible signal fraction of 10\%~\cite{Shi:2019wzi}.
\ampt\ simulations suggest that the final-state partonic and hadronic interactions can significantly reduce the charge separation signal, by an order of magnitude~\cite{Ma:2011uma}.  
The final-state reduction is shown to be very similar between isobar collisions because of the nearly identical conditions created in those  collisions so that the initial-state difference in the CME would persist to the final state, modulo with smaller amplitude~\cite{Deng:2018dut}.

The main objective of this approach is to measure the double ratio between the two isobar systems, $(\dgamma/v_2)_{\Ru}/(\dgamma/v_2)_{\Zr}$.  
Note that the measurement of the ratio $\dgv$ does not require knowledge of the RP resolution~\cite{Voloshin:2018qsm}, which reduces the systematic uncertainty of the result.  
It also ``normalizes'' the $\gamma$ correlator to the elliptic flow value (to which the background is proportional) and thus can be used for a signal comparison in isobar collisions even if the elliptic flow values are slightly different in the two systems.  
Thus, the double ratio $(\Delta\gamma/v_2)_{\rm Ru+Ru}/(\Delta\gamma/v_2)_{\rm Zr+Zr}$, and specifically its deviation from unity, can be used for the detection of the CME signal.  
To extract the CME fraction in this approach the double ratio is fit with the equation:
\begin{equation} 
\frac{\dgv_{\Ru}}{\dgv_{\Zr}}=1+\fcme^{\sms
  \Zr}\left[ \left(\frac{B_{\sms \Ru}}{B_{\sms \Zr}}\right)^2-1\right], 
\label{eq:dblratio}
\end{equation}
where $f_{\textsc{cme}}^{\Zr}$ is the CME fraction in the $\dgamma$ correlator measured in \Zr\ collisions, and $B_{\sms \Ru}/B_{\sms \Zr}$ is the ratio of the magnetic field strengths in \Ru\ and \Zr\ collisions. 
By default, this ratio is taken as the ratio of the nuclear charges, but it can be varied to take into account the uncertainties related to the magnetic field determination.

For a non-zero CME signal, it is expected that the double ratio $(\dgamma/v_2)_{\Ru}/(\dgamma/v_2)_{\Zr}$ would be greater than unity, as the CME signal in \Ru\ collisions is expected to be about 15\% larger than in \Zr\ collisions~\cite{Voloshin:2010ut,Deng:2016knn}.
However, the experimental results did not align with this expectation, indicating a non-negligible background difference between the isobars, which will be discussed in detail in Sect.~\ref{sec:new:isobar}.

\subsubsection{Beam-energy dependence}
\label{sec:methods:energy}
The CME-induced electric current is proportional to the axial charge density and the magnetic field strength; see Eq.~(\ref{eq:j5}). 
The magnetic field peak strength in heavy-ion collisions increases with increasing beam energy, but its lifetime decreases~\cite{Skokov:2009qp,Voronyuk:2011jd,Deng:2012pc}.
The question of magnetic field time evolution is not quantitatively settled because of the lack of knowledge on the electric conductivity of the QGP~\cite{Tuchin:2013apa,McLerran:2013hla,Li:2018ufq,Huang:2022qdn}.
The axial charge density is expected to be larger at higher beam energies and could disappear at low energies where quarks and gluons are not deconfined. 
It is therefore generally expected that there exist a sweat spot in beam energy to look for the CME, however, the question of its location will have to be answered experimentally. 
Thus, it remains viable to analyze all available heavy-ion collision data, such as those from the RHIC beam energy scan program.

\subsubsection{Event-shape engineering}
\label{sec:methods:ESE}

The main background to the CME measurements is the flow-induced charge-dependent correlations of Eq.~(\ref{eq:cluster}), which are proportional to the elliptic flow $v_2$.  
Due to the initial geometry fluctuations, the elliptic flow also fluctuates event-by-event. 
One may therefore try to select the events with large or small elliptic flow (and correspondingly the background) while keeping the possible CME signal, determined mostly by the number of spectators, unchanged.  
This can be achieved through the so-called event-shape engineering (ESE)~\cite{Schukraft:2012ah}.  
In this approach, the event selection is based on the flow vector
\begin{equation}\label{eq:q}
    \mathbf{q}_2=\sqrt{N}\left(\mean{\cos2\phi},\mean{\sin2\phi}\right),
\end{equation}
calculated with particles in a given momentum region (usually at forward or backward rapidities) for events within a narrow centrality bin (so that the CME signal is approximately the same).  
It was shown that the magnitude, $q_2$, is closely related to the average elliptic flow $v_2$ in the event.  
The latter is evaluated by calculating~$v_2$ in another momentum window (usually at midrapidity) to avoid autocorrelation between $q_2$ and flow measurements, making statistical fluctuations in $q_2$ and estimates of flow independent.
One then studies the physics of interest (in our case the CME) as a function of the event $v_2$.  
The results obtained with this method are presented in Sect.~\ref{sec:new:ESE}.

In midcentral collisions, the ESE technique allows one to select events that can differ in $v_2$ by almost a factor of two~\cite{Acharya:2017fau}. 
However, the probability of events with small/large $v_2$ extremes is low, which is reflected in the statistical errors. 
For the estimate of the CME fraction, one needs to extrapolate such a dependence toward zero $v_2$ values, which can be statistically limited. 

On the other hand, the statistical fluctuations in the apparent event-by-event anisotropy quantities (e.g.~the $v_2^\obs\equiv \mean{\cos(2\phi-2\psiep)}$, where the average is taken over particles in a given event~\cite{STAR:2013zgu}) are large, and it is tempting to use them to select events where the background contribution is small due not only to the small(er) ellipticity of the initial geometry but also to statistical fluctuations in particle distribution in a given event. 
The so-called event-shape selection method (ESS) is based on this idea. 
Because the selection of events in this approach is based on statistical fluctuations, or {\em apparent} flow, precise calculations of the corresponding biases are very difficult, which, from our point of view, makes the interpretation of the results obtained in this way impractical. 
The discussion of the results obtained with this method as well as more recent development of the method is presented in Sect.~\ref{sec:results:ESS}

\subsubsection{Measurements relative to  spectator and participant planes}
\label{sec:methods:ppsp}

Measurements relative to the second harmonic participant plane yield by definition the largest values of elliptic flow. 
The elliptic flow values measured with respect to the participant plane, $\vtwopp$, and to the spectator plane, $\vtwosp$, 
differ by 10--20\% depending on the collision centrality.  
The CME-induced charge separation is the largest in the direction of the magnetic field, determined primarily by spectator protons. 
It is plausible to assume that the magnetic field is on average perpendicular to the spectator plane.  
Therefore, the CME contribution to $\dgamma$ would be the largest if measured with respect to the spectator plane and would be reduced in magnitude if measured relative to the participant plane.  
The flow-induced background, on the other hand, is proportional to elliptic flow that is strongest in the participant plane.
The relative magnitudes of $v_2$ and CME signals in two cases are likely reciprocal, i.e.,
\begin{equation}
  \vtwosp/\vtwopp = \dg_\cme^\spp/\dg_\cme^\ssp =
  \mean{\cos2(\psi_\spp-\psi_\ssp)}\equiv a.
\end{equation}
These features can be used to extract the CME signal fraction from the two $\dg$ measurements with respect to SP and PP, in place of $\psi_\srp$ in Eq.~(\ref{eq:gamma}), by~\cite{Xu:2017qfs,Voloshin:2018qsm}
\begin{equation}\label{eq:fcme_obs}
    \fcme^\spp\equiv\frac{\dg_\cme^\spp}{\dg^\spp}=\frac{A/a-1}{1/a^2-1}\,,
\end{equation}
where $A=\dgsp/\dgpp$.  

Note that the calculation of the double ratio, $(\dgsp/\vtwosp)/(\dgpp/\vtwopp)=A/a$, does not require knowledge of the RP resolutions and can be measured more accurately than, e.g., the ratio $\vtwosp/\vtwopp=a$. 
Then the deviation of the double ratio from unity would immediately indicate the presence of the CME contribution, as can be seen from the equation below
\begin{equation}
  \frac{\dgv^\ssp}{\dgv^\spp} =
  1 + \fcme^\spp \left[ \left( \frac{\vtwopp}{\vtwosp}\right)^2 - 1
\right]\,.
\label{eq:Group4_fcmeSP}
\end{equation}

The flow-induced background is identified by design in this SP/PP
method~\cite{Xu:2017qfs,Voloshin:2018qsm}. 
The method is unique in the sense that it does not depend on the specific details of the physics background, whether it is induced by collective flow or, for example, by spin alignment of vector mesons from color field fluctuations;
as long as the physics background contributing to $\dg$ is proportional to flow, it is accounted for by the method.

The SP/PP method is similar in spirit to the idea of CME detection in isobar collisions -- both approaches  compare two measurements that supposedly differ in $\fcme$. 
The SP/PP method has an advantage in that the collision events used for SP and PP measurements are identical, whereas the background in the isobar collisions might have subtle differences, due to differences in the centrality selections and/or isobar nuclear structures.

%% file: early.tex
\section{Early Measurements}
\label{sec:early}

\subsection{First results on the gamma correlator}
\label{sec:first}
 
\begin{figure}
  \includegraphics[width=0.95\linewidth]{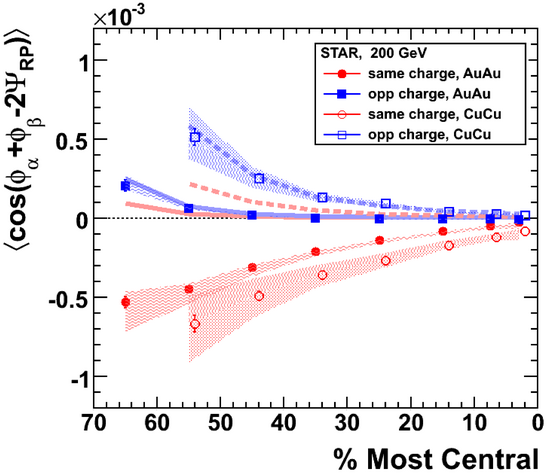}
  \caption{\label{fig:STAR2009}(Color online) The first measurements
    of the opposite-sign (OS) and same-sign (SS) $\gamma$
    correlators in Au+Au and Cu+Cu collisions at
    $\snn=200$~GeV by STAR~\cite{STAR:2009wot,STAR:2009tro}.
    The thick solid (Au+Au) and dashed (Cu+Cu) lines
    represent \hijing\ calculations presenting the contributions from
    the RP-independent three-particle correlations. Shaded bands represent
    uncertainty from the measurement of $v_2$. 
    Figure is taken from Refs.~\cite{STAR:2009wot,STAR:2009tro}.}
\end{figure}

The first results on the CME search in Au+Au and Cu+Cu collisions at $\snn=200$~GeV were published by the STAR Collaboration in 2009~\cite{STAR:2009wot,STAR:2009tro}. As shown in Fig.~\ref{fig:STAR2009}, a clear difference was observed between the correlators $\gos$ and $\gss$.  
The results were consistent with the general expectations for the CME signal on top of a common negative background, which could be attributed to the charge-independent correlations.

\begin{figure}
  \includegraphics[width=0.9\linewidth]{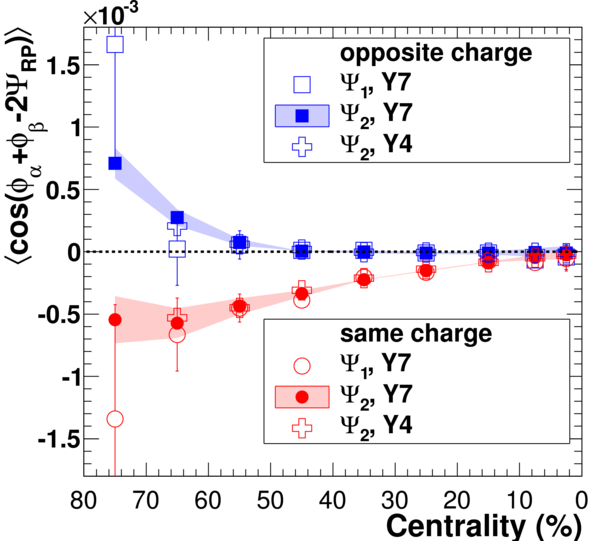}
  \caption{\label{fig:STAR_ZDC}(Color online) 
    Three-point $\gos$ and $\gss$ correlators versus centrality in Au+Au collisions at  $\snn=200$~GeV measured by STAR with the first-order harmonic plane determined by the ZDCs~\cite{STAR:2013ksd}, compared to those measured in    Ref.~\cite{STAR:2009wot,STAR:2009tro} with the second-order harmonic plane from the TPC. 
    Shaded areas for the second harmonic results represent the systematic uncertainty of the event plane determination. Figure is taken from Ref.~\cite{STAR:2013ksd}.
  }
\end{figure}

STAR has also measured the $\gamma$ correlators with respect to the first-order harmonic plane $\psi_1$ from the zero-degree calorimeters (ZDCs) determined by the spectator neutrons~\cite{STAR:2013ksd}.  
The results, cf.~Fig.~\ref{fig:STAR_ZDC}, were consistent with the measurements with respect to the Time Projection Chamber (TPC) $\psi_2$. 
As discussed in Sect.~\ref{sec:methods:ppsp}, the TPC (PP) and ZDC (SP) measurements are expected to differ, however, the statistics were not sufficient to reveal the difference.

\begin{figure}
  \includegraphics[width=0.95\linewidth]{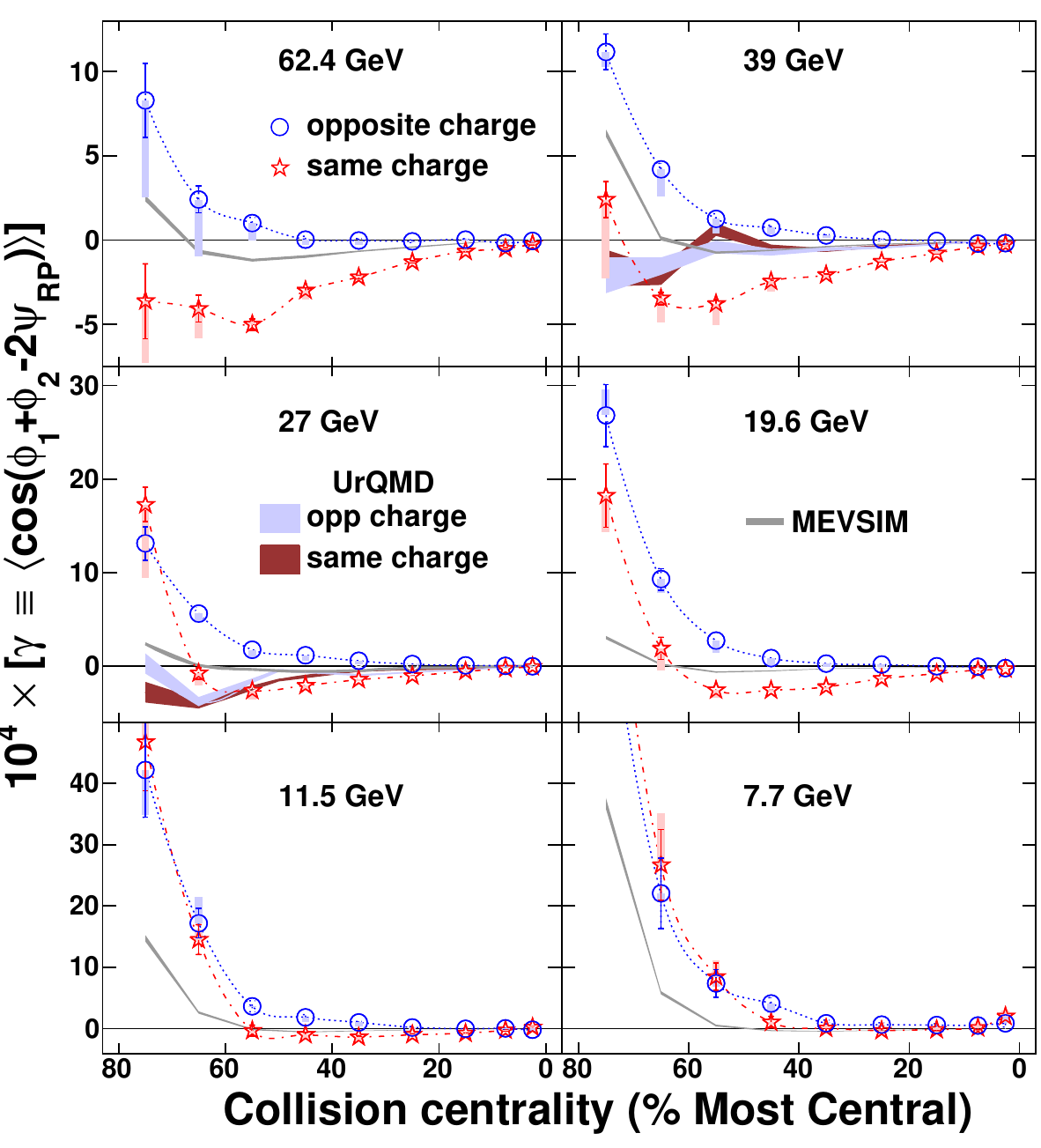}
  \caption{\label{fig:STAR_BES}(Color online) 
    The three-point correlators $\gos$ and $\gss$ as a function of centrality for Au+Au collisions at $\snn=7.7$-–62.4~GeV~\cite{STAR:2014uiw} from STAR. 
    Charge-independent results from calculations of the MEVSIM event generator~\cite{Ray:2000se} are shown as gray curves. 
    The $\gos$ and $\gss$ from UrQMD calculations are  shown as shaded bands for 27 and 39~GeV data points. 
    Note that the vertical scales are different for different rows. 
    Figure is taken from Ref.~\cite{STAR:2014uiw}.}
\end{figure}

STAR has also measured the $\gamma$ correlators at lower RHIC energies from
the Beam Energy Scan phase-I (BES-I) data~\cite{STAR:2014uiw}.  
These results are shown in Fig.~\ref{fig:STAR_BES}. 
Differences between $\gos$ and $\gss$ were observed at all collision energies except at the two lowest energies of 11.5 and 7.7~GeV.  
This could suggest that the CME disappears at these energies, consistent with the expectations that the chiral symmetry restoration needed for the CME does not occur at temperatures achievable in the collisions at low energies.

\begin{figure}
  \includegraphics[width=\linewidth]{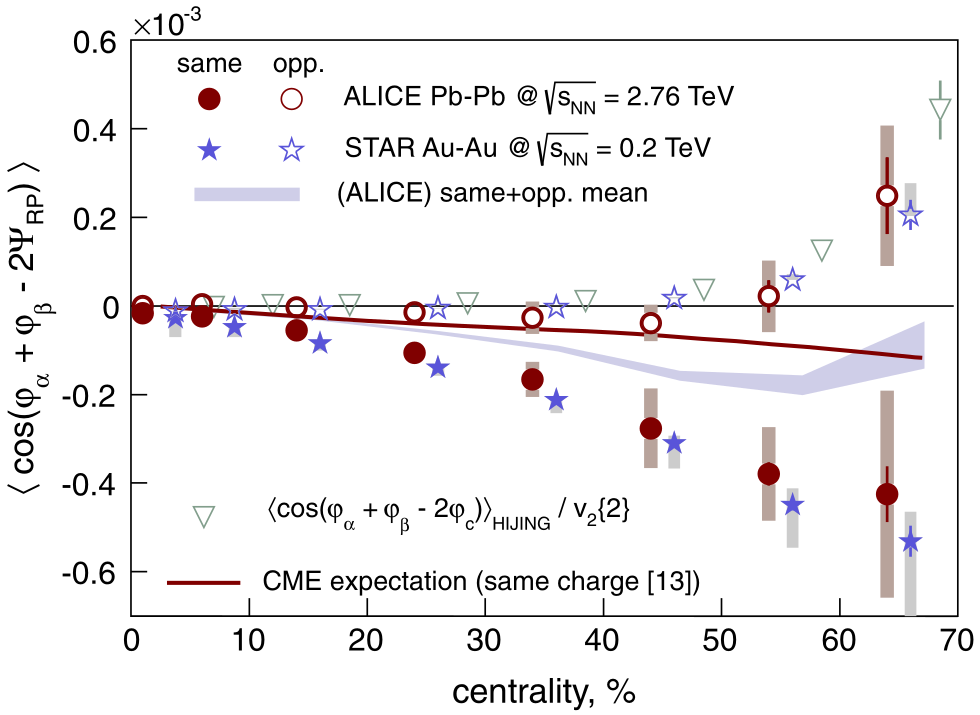}
  \caption{\label{fig:ALICE2013}(Color online) 
    The centrality dependence of $\gamma$ correlators.
    The circles indicate the ALICE results~\cite{Abelev:2012pa} and the stars show the STAR data~\cite{STAR:2009tro,STAR:2009wot}. 
    The triangles represent the three-particle correlations, Eq.~\ref{eq:c3}, from   \hijing\ corrected for the experimentally measured $v_2\two$. 
    Points are displaced horizontally for visibility. 
    See Ref.~\cite{Abelev:2012pa} for more details.
    Figure is taken from Ref.~\cite{Abelev:2012pa}.}
\end{figure}

The ALICE experiment at the LHC measured the $\gamma$ correlator in Pb+Pb collisions at $\snn=2.76$~TeV~\cite{Abelev:2012pa}.  
The results, shown in Fig.~\ref{fig:ALICE2013}, exhibit similar trends and are close in magnitude compared to the STAR measurements at RHIC.  
While similar trends are generally expected, the similar magnitudes are surprising, given the differences in the average multiplicities, lifetimes of the magnetic field, values of the elliptic flow, and somewhat different acceptances.

\begin{figure}
  \includegraphics[width=\linewidth]{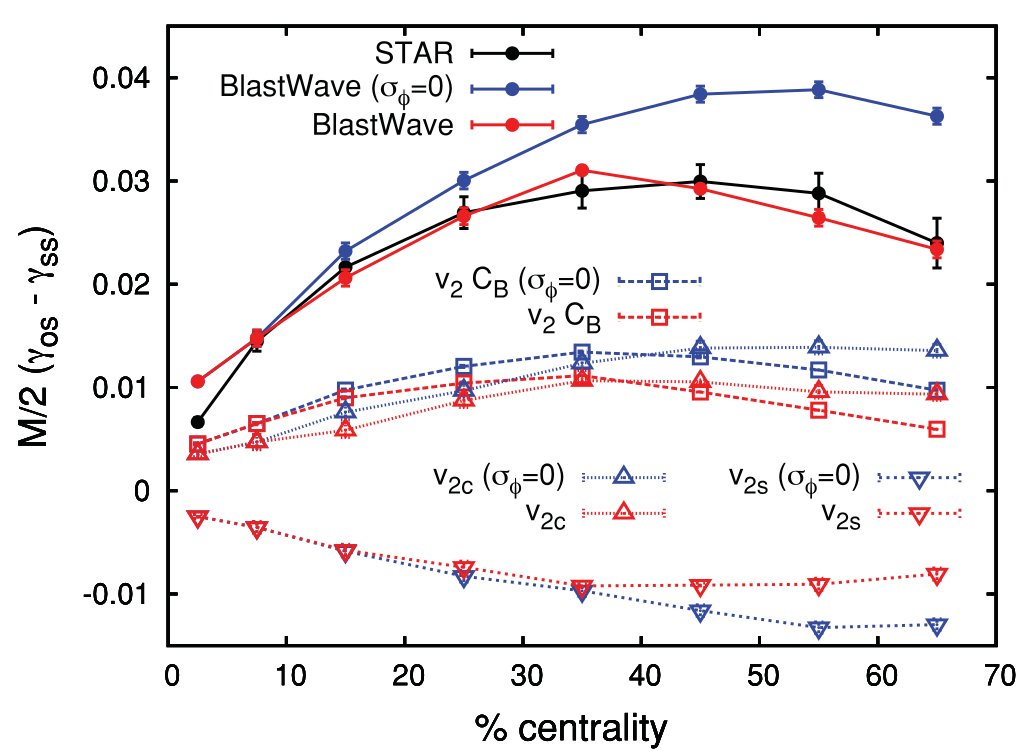}
  \caption{\label{fig:BW}(Color online) The multiplicity-scaled $\dg$
    from Blast-Wave model calculations~\cite{Schlichting:2010qia}
    for realistic charge separation at freeze-out (red dots) and
    perfectly local charge conservation (blue dots), compared to the STAR
    data (black dots). For explanation of the dashed curves, presenting
    separate contributions, see Ref.~\cite{Schlichting:2010qia}. Figure is taken from Ref.~\cite{Schlichting:2010qia}.}
\end{figure}

The transport model calculation of the charge-dependent background performed by STAR~\cite{STAR:2009tro,STAR:2009wot} 
could explain only about 1/3 of the observed signal.  
On the other hand, the Blast-Wave (BW) calculations~\cite{Schlichting:2010qia}, including the LCC and based on the parameterization of the momentum spectra and elliptic anisotropy data of Au+Au collisions at RHIC, with reasonable parameters, can reproduce the STAR measurements; see Fig.~\ref{fig:BW}.
Similar BW+LCC calculations~\cite{Wu:2022fwz} at LHC energies can simultaneously describe both the CME and CMW measurements by the ALICE Collaboration.

The LCC picture adopted in~\cite{Schlichting:2010qia,Wu:2022fwz} assumes that the pairs of opposite charges are created very close in space at the late stage of the system evolution. 
Radial boost of the pair due to transverse expansion leads to particle collimation in azimuth and pseudorapidity~\cite{Voloshin:2003ud,Voloshin:2004th}. 
Due to elliptic flow, the collimation is stronger in-plane than out-of-plane, which contributes to $\dgamma$~\cite{Schlichting:2010qia}.
While this mechanism is probably the strongest candidate for the background description, we note that the parameters of the BW model in~\cite{Schlichting:2010qia} were {\em tuned} to the measured charge balance function, the detailed shape of which by itself includes the possible CME.

\subsection{Higher harmonics. The kappa parameter}
\label{sec:first:kappa}

\begin{figure*}[htbp]
\includegraphics[width=0.95\linewidth]{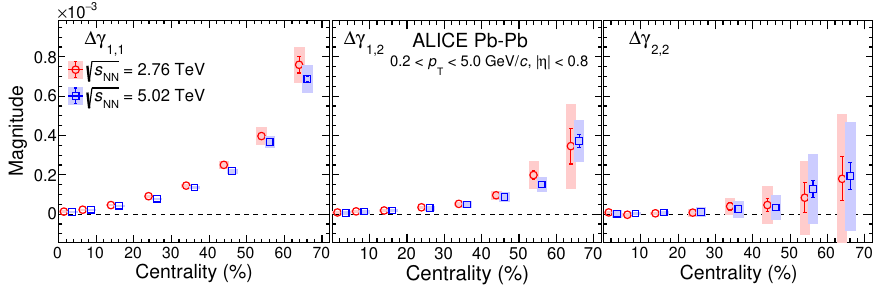}
\vspace*{-4mm}
\caption{(Color online)
    Comparison of the charge-dependent parts of correlators
$\gamma_{1,1}\equiv\mean{\cos(\phi_\alpha+\phi_\beta-2\psi_2)}$,
$\gamma_{1,2}\equiv\mean{\cos(\phi_\alpha+2\phi_\beta-3\psi_3)}$, and $\gamma_{2,2}\equiv\mean{\cos(2\phi_\alpha+2\phi_\beta-4\psi_4)}$  as a function of centrality at the LHC energies.  Figure is taken from Ref.~\cite{ALICE:2020siw}.}
\label{fig:jm}
\end{figure*}

As discussed in Section~\ref{sec:methods:mixed} the measurements including higher harmonics, free from the CME contribution, could clarify the origin of the charge-dependent correlations and the role of the LCC.
The correlations measured with respect to the fourth harmonic plane, $\mean{\cos(2\phi_\alpha+2\phi_\beta-4\psi_4)}$, should not contain any contribution from the CME, while it should include the effect of the LCC~\cite{Voloshin:2011mx,Voloshin:2012fv}.
The correlations due to the LCC in this case are expected to be somewhat smaller in magnitude as the fourth-harmonic flow is not as strong as the elliptic flow. 
The ALICE results including these measurements are presented in Fig.~\ref{fig:jm}~\cite{ALICE:2020siw}. 
The correlations relative to the fourth harmonic appear to be very small, but  detailed model calculations would be needed to draw any definite conclusion from this measurement. 
The accuracy of such calculations at the moment is not good enough for the purpose of identifying the CME.
The measurements including mixed harmonics have also been performed by the CMS and ALICE Collaboration~\cite{CMS:2017pah,ALICE:2020siw}, but, as noticed in Section~\ref{sec:methods:mixed}, the interpretation of those measurements are rather complicated and requires by itself a detailed  knowledge of the background correlations.

Following the proposal~\cite{Bzdak:2012ia} to relate the $\dg$ and $\dd$ measurements, Eqs.~(\ref{eq:delta}--\ref{eq:HF}), in order to estimate the  background, the STAR Collaboration analyzed the data with the equations~\cite{STAR:2014uiw}:
\begin{subequations}\label{eq:kappa}
    \begin{align}   
        \dg &= \kappa v_2 F - H \,, \label{eq:FHkappa}\\
        \dd &= F + H \,,
    \end{align}
\end{subequations}
where a fudge factor $\kappa$ was introduced in attempt to cover the approximations made in the steps of Eq.~(\ref{eq:gamma-delta}).
Note that the value of the $\kappa$ parameter cannot be theoretically calculated or experimentally measured, and using {\em ad-hoc} guessed values of $\kappa$ to extract a CME signal is not rigorous or accurate.

Comparing  Eqs.~\ref{eq:kappa} to the background contribution given in Eq.~(\ref{eq:cluster}), the $\kappa$ parameter is related to the azimuathal distribution of cluster decay products as
\begin{equation}
  \label{eq:kappa2}
  \kappa = \frac{\mean{\cos(\phi_\alpha+\phi_\beta-2\phi_{\rm cl})}}
         {\mean{\cos(\phi_\alpha-\phi_\beta)}} \cdot \frac{v_{2,{\rm cl}}}{v_2}\,.
\end{equation}
From this relationship, one can expect that the values of $\kappa$ to  be of the order of unity. 
This is approximately what has been measured by the CMS Collaboration~\cite{CMS:2017lrw} by
\begin{equation}
    \kappa = \frac{\dg}{v_2\dd}\,.
\end{equation}
Unfortunately, the large uncertainties in the actual $\kappa$ values do not allow for any quantitative measurement of the CME signal.

\subsection{Measurements in small systems}
\label{sec:small}

\begin{figure}
  \includegraphics[width=0.9\linewidth]{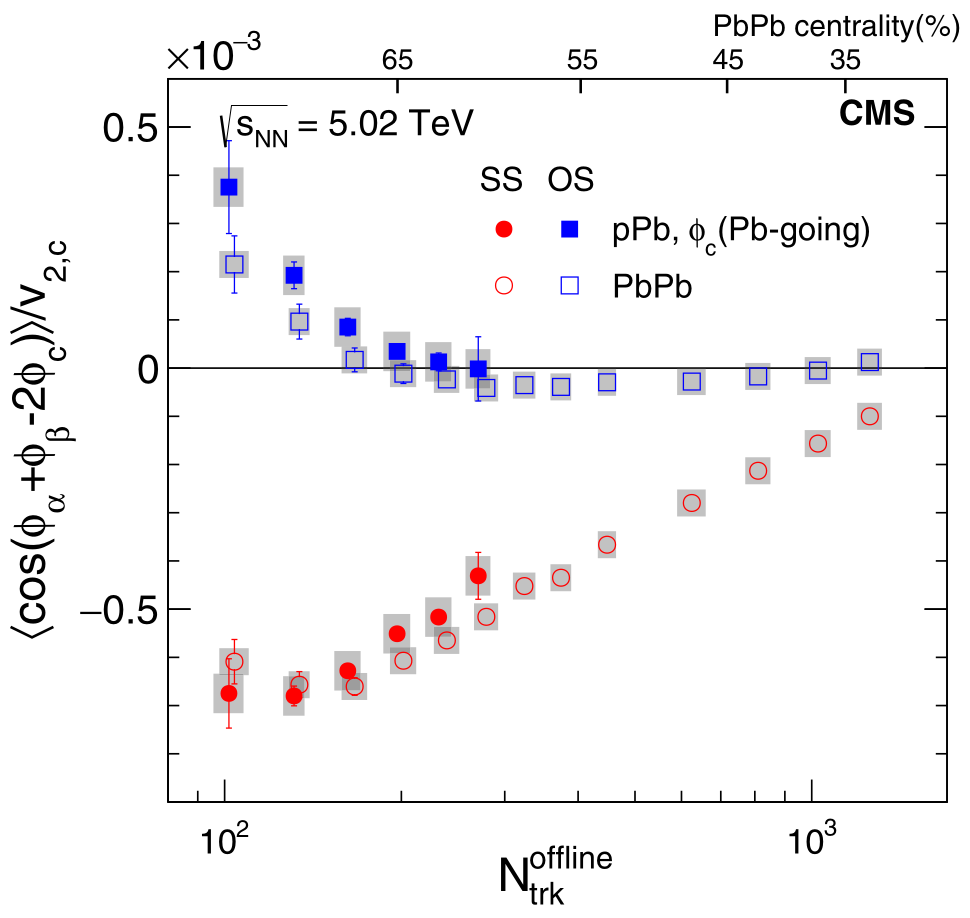}
  \includegraphics[width=0.88\linewidth]{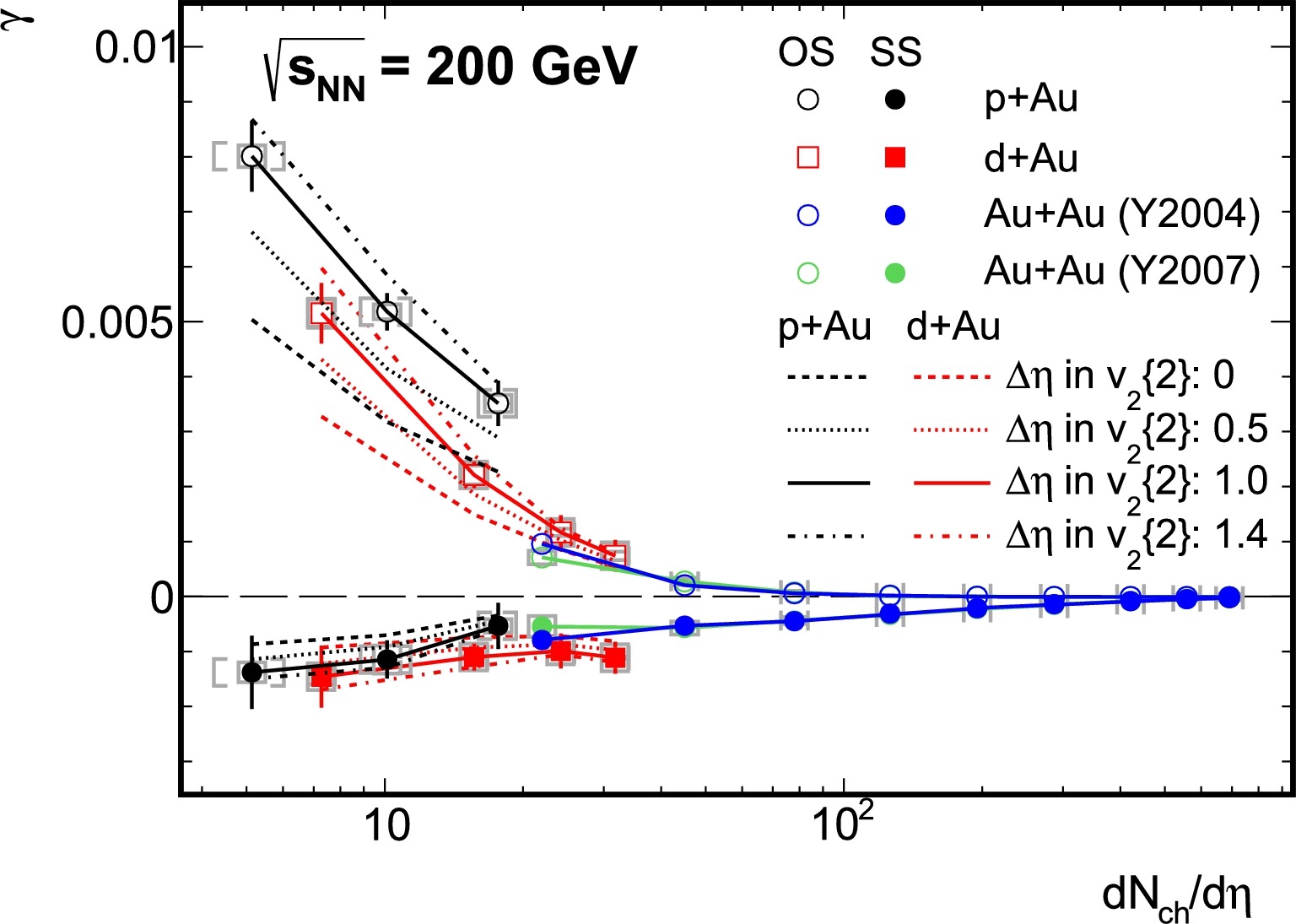}
  \caption{\label{fig:small}(Color online) 
    The $\gos$ and $\gss$ correlators in small systems compared to those in large systems as functions of multiplicity. 
    Upper plot shows the CMS results in $p$+Pb and Pb+Pb collisions at $\snn=5.02$~TeV~\cite{CMS:2016wfo}. 
    The results are averaged over $|\eta_\alpha-\eta_\beta|<1.6$. 
    Statistical and systematic uncertainties are indicated by the error bars and shaded regions, respectively. 
    Plot is taken from Ref.~\cite{CMS:2016wfo}. 
    Lower plot shows STAR results in $p$+Au, $d$+Au, and Au+Au  collisions~\cite{STAR:2019xzd}. 
    Particles $\alpha,\beta$, and $c$ are all from the full TPC acceptance $|\eta|<1$; no $\eta$ gap is applied.
    Statistical uncertainties are shown by the vertical bars and systematic ones are shown by the vertical brackets. 
    The horizontal brackets indicate the systematic uncertainty in $d\Nch/d\eta$. 
    Plot is taken from Ref.~\cite{STAR:2019xzd}.}
\end{figure}

Another approach proposed to assess the magnitude of the background in the CME search is the measurements in small systems, where the CME signal is expected to be negligible. 
Such measurements have been performed in p+Pb collisions by the CMS Collaboration at the LHC~\cite{CMS:2016wfo}, and in $p$+Au and $d$+Au collisions by the STAR Collaboration at RHIC~\cite{STAR:2019xzd}. 
The CMS and STAR small-system results are shown in Fig.~\ref{fig:small}, where large differences between $\gos$ and $\gss$ are observed, comparable to  those measured in peripheral heavy-ion collisions.  
Because the magnetic field direction is not correlated with the participant plane direction in small systems, which arises from pure fluctuation effects, any CME signals, even existent in those small-system collisions, are not observable by the $\dg$ correlator~\cite{CMS:2016wfo,Belmont:2016oqp}.
Thus, the observed $\dg$ in small systems could indicate similarly large background contributions in the heavy-ion measurements.

Unfortunately, the measurements of the $\dg$ correlator in small-system collisions can not be used to  assess the background in the large systems, as they are completely dominated by the RP-independent three-particle correlations (see Sect.~\ref{sec:RPindependent}).
This conclusion was already reached in the very first STAR papers~\cite{STAR:2009wot,STAR:2009tro}, in which it was shown  that the measurements in peripheral collisions can be described by the RP-independent background.
More recent \hijing~model calculations~\cite{Zhao:2019kyk} confirm that the $\gamma$ correlators in peripheral collisions at RHIC energies can be quantitatively described in this model, suggesting the RP-independent three-particle correlation nature of the measured $\gamma$ correlators.  
Therefore, an apparent quantitative similarity of the $\gamma$ correlator as a function of multiplicity in small systems and peripheral heavy-ion collisions is not at all surprising, as it comes from the same physics.

\subsection{Gamma correlator vs pair invariant mass}
\label{sec:minv}

\begin{figure}
  \includegraphics[width=\linewidth]{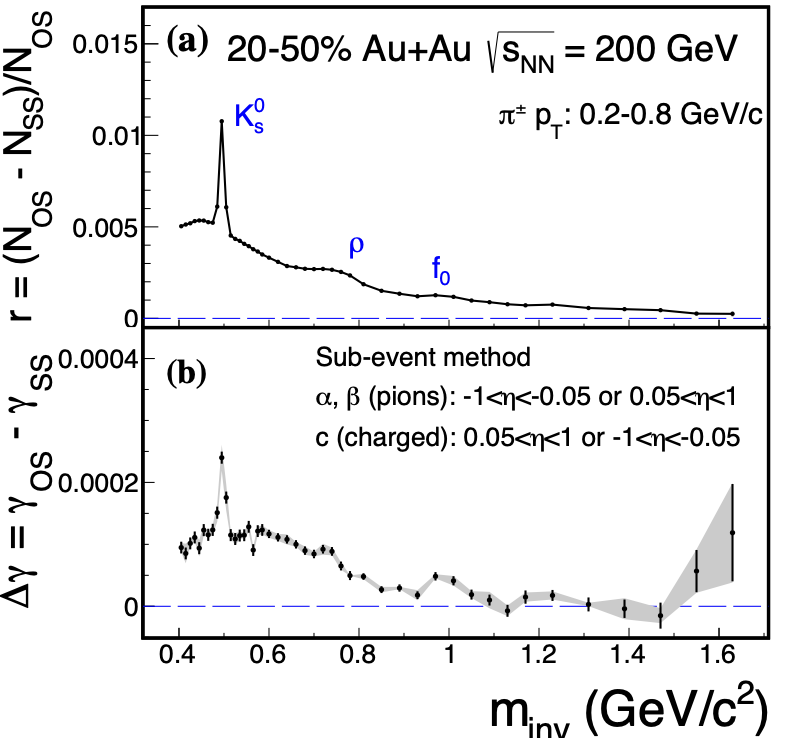}
  \caption{ (Color online) 
    (a) The relative excess of OS over SS pion
    pairs, and (b) $\dg$ in 20--50\% Au+Au collisions at $\snn=200$~GeV
    as functions of the pair invariant mass measured by STAR. 
    Error bars are statistical. 
    The shaded area in panel (b) shows systematic uncertainties. 
    Figure is taken from Ref.~\cite{STAR:2020gky}.}
  \label{fig:STAR_minv}
\end{figure}

More insight into the nature of background correlation sources could be obtained with the $\dg$ measurement as a function of the POI pair invariant mass ($\minv$)~\cite{Zhao:2017nfq,STAR:2020gky}. 
Figure~\ref{fig:STAR_minv}, upper panel, presents the relative difference in the OS and SS pair multiplicities as a function of $\minv$, to be compared to the $\dg$ correlator shown in the lower panel.  
The $K_S^0$, $\rho^0$, and $f_0$ resonance peaks are apparent in the $\minv$ spectrum;  the continuum underneath the resonance peaks is suggestive of the LCC type correlations.
The $\dg$ correlator also exhibits peaks at resonance mass locations, indicating a significant background contribution of these resonances to $\dg$.
The interpretation of the measurements in the continuum regions is more complicated.
Note that the clusters of particles contributing to the CME are also expected to contribute to the excess of OS pairs over SS pairs in the region $\minv \lesssim 1$~GeV, the typical scale of the instanton/sphaleron clusters.

\subsection{Event-shape selection}
\label{sec:results:ESS}

\begin{figure}[hbt]
  \includegraphics[width=0.85\linewidth]{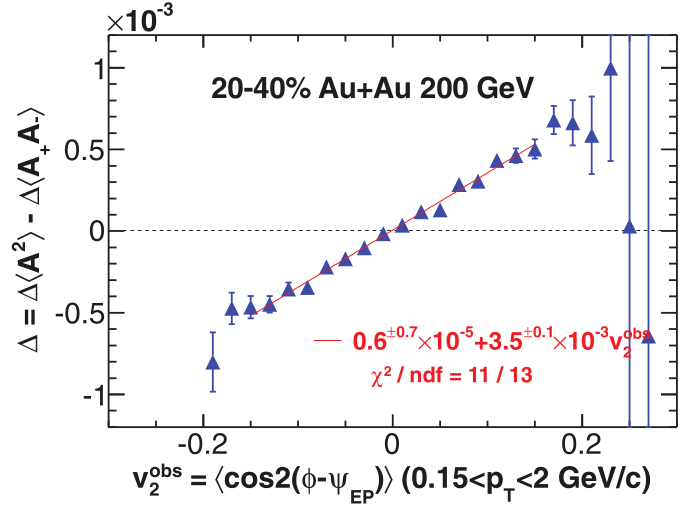}
  \caption{\label{fig:STAR_ESS1}(Color online) 
    Charge  multiplicity asymmetry correlations  $\Delta\mean{A^2}-\Delta\mean{A_+A_-}$~\cite{STAR:2013zgu} (a quantity similar to $\dg$) as a function of $v_2^\obs$ in 20\%–40\% central Au+Au collisions at $\snn=200$~GeV from STAR. Both quantities are measured in the STAR TPC with respect to the first-order harmonic plane reconstructed from the ZDC shower maximum detectors. 
    Figure is taken from Ref.~\cite{STAR:2013zgu}.
    }
\end{figure}

The first application of the event-shape selection technique (see Sect.~\ref{sec:methods:ESE}) was reported by STAR Collaboration in Ref.~\cite{STAR:2013zgu}.
In this analysis, the events were selected according to the values of $v_2^\obs=\mean{\cos2(\phi_\poi-\psi_{\rm EP})}$ calculated for POIs relative to the EP reconstructed from other particles. 
Because of the statistical fluctuations due to the finite number of particles used in the calculation, $v_2^\obs$ varies over a wide range including $v_2^\obs=0$ and extends even to the negative $v_2^\obs$ values.
Figure~\ref{fig:STAR_ESS1} presents an example of the dependence on $v_2^\obs$ of a quantity similar to $\dg$~\cite{STAR:2013zgu}. 
The intercept of a linear fit is well constrained by the data points.  
Of course, the events with $v_2^\obs=0$ are strongly {\em biased} towards a ``round'' emission envelope of POIs in terms of the reconstructed EP, and one would be tempting to conclude that the background to the CME is also zero in these events.
Unfortunately, this is not the case: When the events are binned according to the event-shape quantity $v_2^\obs$, the average $v_2$ of the ``flowing clusters'' is not zero but positive. This has been verified by a toy-model simulation in Ref.~\cite{Zhao:2019kyk}. 
The RP-dependent background can only partially be suppressed in this method, but how much background remains is practically unknown.

\begin{figure}[hbt]
  \includegraphics[width=1.\linewidth]{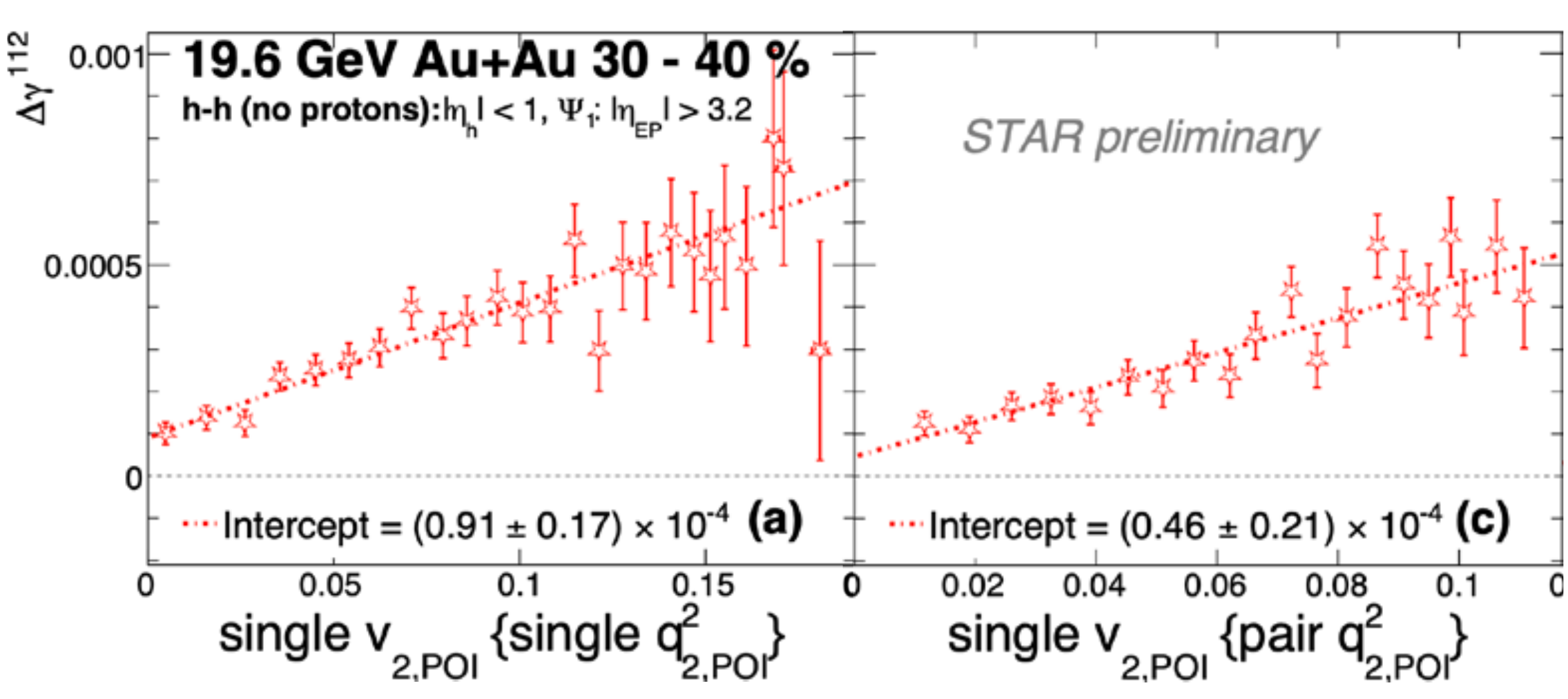}
  \caption{(Color online)  
  The $\dg$ correlator as a function of single particle $v_2$ of POIs (excluding protons and antiprotons) in the 30–40\% centrality range of Au+Au collisions at 19.6~GeV measured by STAR.    
  The  ESS is performed based on $q_2^2$ calculated using either single particles or particle pairs of POIs. 
  The POIs are taken from the $|\eta| <1$ region, and the measurements are performed using the spectator plane from EPD ($|\eta| > 3.2$).
  Figure is taken  from  Ref.~\cite{Xu:2023wcy}.}
  \label{fig:STAR_ESS2}
\end{figure}

More recently, it has been proposed that the $q_2^2$ of pairs of POIs be used for the event-shape selection.  
Figure~\ref{fig:STAR_ESS2} shows the preliminary results from STAR Collaboration~\cite{Xu:2023wcy} where the $\dg$ is plotted vs~single particle $v_2$ for the events selected according to ``single'' $q_2$ values or that of particles pairs. 
Unfortunately, the interpretation of the intercept, even its sign, is rather uncertain. 
For example, the \avfd\ model calculations suggest that this method seems to do a good job in eliminating backgrounds allowing extraction of the correct CME magnitude implemented in the model, whereas it is shown that the method gives a negative intercept for \ampt\ events where no CME is present~\cite{Xu:2023elq}.  
It is further shown in Ref.~\cite{Li:2024gdz} that the method yields positive intercepts for the \epos\ and \hydjet\ models, in the absence of any CME signals; the results also depend on the collision energies.

%% file: recent.tex
\section{Experimental Searches with Improved Techniques}
\label{sec:new}

\subsection{Event-shape engineering}
\label{sec:new:ESE}

\begin{figure}
  \includegraphics[width=0.95\linewidth]{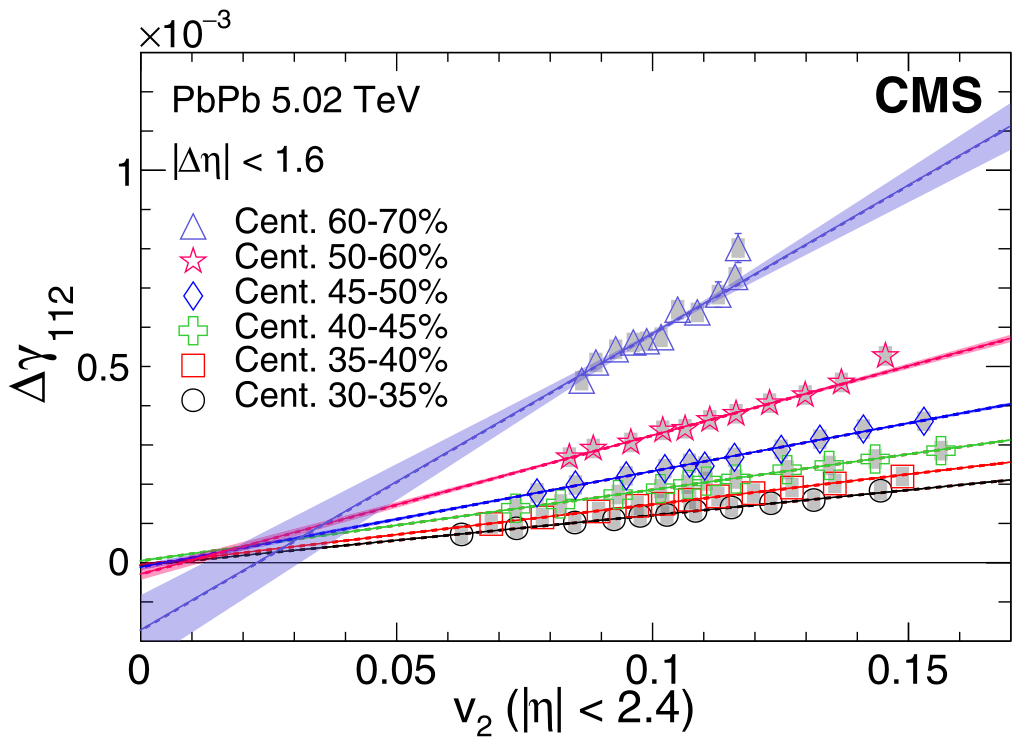}
  \caption{\label{fig:CMS_ESE}(Color online) 
  The $\dg_{112}\equiv\dg$  averaged over $|\deta| < 1.6$ as a function of $v_2$ evaluated in    each $q_2$ class by CMS.  
  Statistical (mostly smaller than the symbol  size) and systematic uncertainties are indicated by the error bars  and shaded regions, respectively. 
  A one standard deviation  uncertainty from the fit is also show. 
  Figure is taken from  Ref.~\cite{CMS:2017lrw}.}
\end{figure}

The CMS experiment~\cite{CMS:2017lrw} applied the ESE method in their CME search in Pb+Pb collisions at $\snn=5.02$~TeV using ${q}_2$ calculated from the forward/backward hadronic calorimeters covering the pseudorapidity rages $4.4 < |\eta| < 5$.  
The main results of this analysis are shown in Fig.~\ref{fig:CMS_ESE}.  
An approximately linear dependence on $v_2$ of the $\dg$ calculated using particles at midrapidity is observed in each centrality bin.
With the current statistical uncertainties, the CME signal (the intercept at $v_2=0$) was found to be consistent with zero~\cite{CMS:2017lrw} with an upper limit of $\fcme <7$\%. 
Note that the CMS Collaboration did not consider that the CME signal might depend on $v_2$ as the correlation of the magnetic field direction to the RP likely decreases with decreasing $v_2$. 
Therefore, the upper limit obtained by CMS is probably underestimated.

\begin{figure}
  \includegraphics[width=0.95\linewidth]{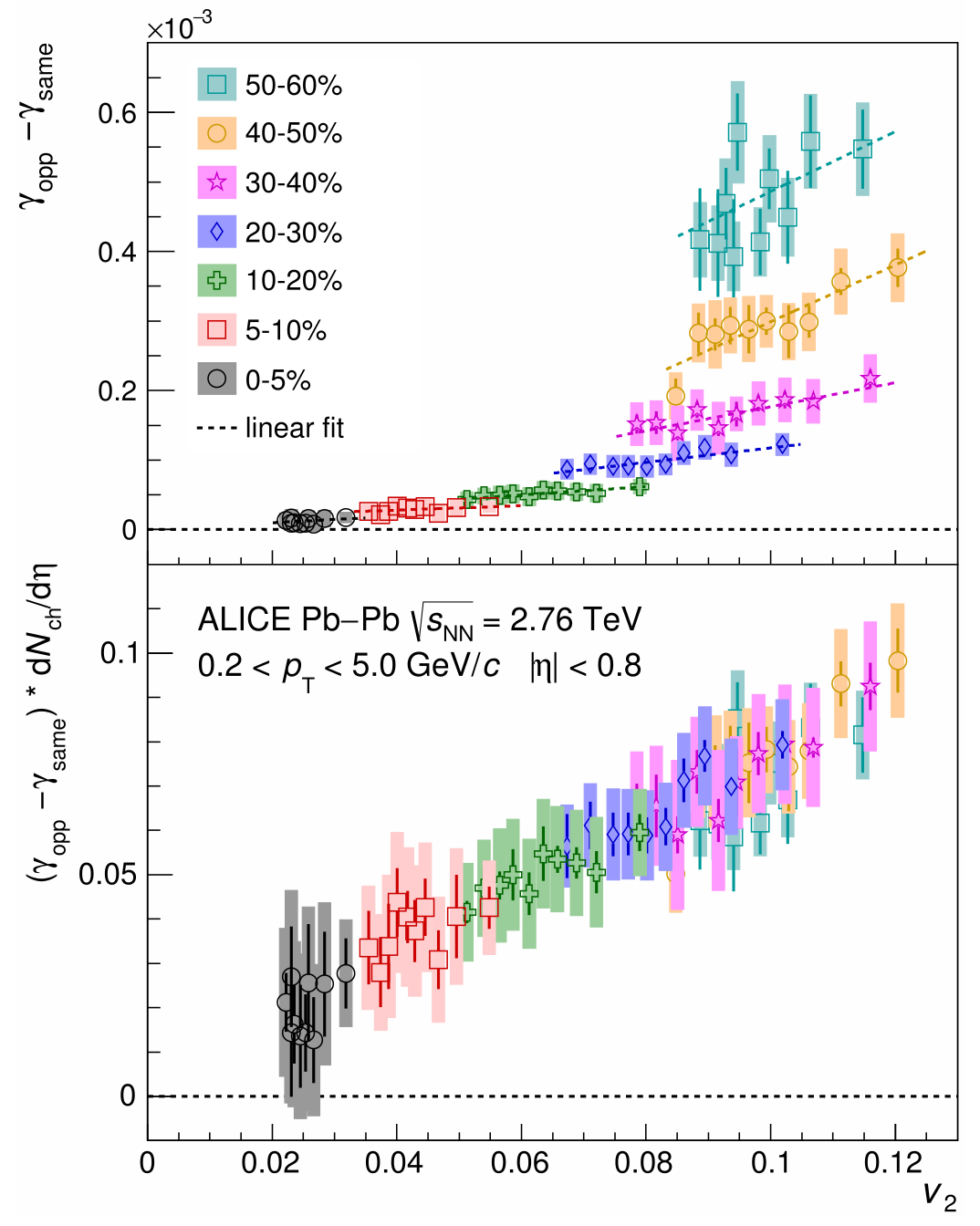}
  \caption{\label{fig:ALICE_ESE}(Color online) 
  The $\dg$ (upper  panel) and the multiplicity-scaled $d\Nch/d\eta \times \dg$ (lower    panel) as functions of $v_2$ for ESE selected events in various centrality classes by ALICE.  
  Error bars (shaded boxes) represent the  statistical (systematic) uncertainties. Dashed lines in the upper    panel are linear fits. 
  Figure is taken from Ref.~\cite{Acharya:2017fau}.}
\end{figure}

The ESE method has also been applied by the ALICE experiment~\cite{Acharya:2017fau}.  Figure~\ref{fig:ALICE_ESE} shows the $\dg$ correlator at midrapidity as a function of $v_2$ in events binned in $q_2^2$ within each centrality bin. 
In this analysis ${q}_2$ is measured using a scintillator array detector (V0C) within pseudorapidity range $-3.7<\eta<-1.7$, while $v_2$ was calculated by the EP method using EPs reconstructed in the V0A detector ($2.8<\eta<5.1$).
The lower panel of Fig.~\ref{fig:ALICE_ESE} shows the multiplicity scaled $\dg$ correlator, which is supposed to mitigate the multiplicity dilution effect.

\begin{figure}
  \includegraphics[width=0.95\linewidth]{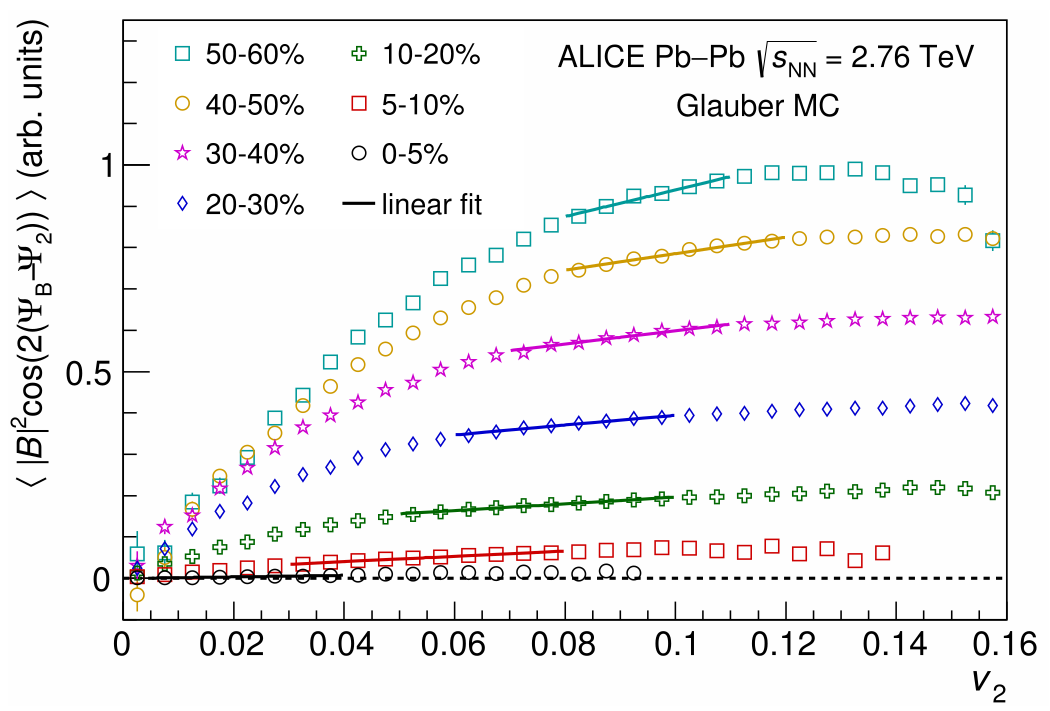}
  \caption{\label{fig:ALICE_ESE_B2}(Color online) 
  The   dependence of the CME signal on $v_2$ for various centrality classes from MC-Glauber simulations by ALICE. 
  No event shape selection is  performed in the model, and therefore a large range in $v_{2}$ is  covered. 
  The solid lines depict linear fits based on the $v_2$  variation observed within each centrality interval. 
  Figure is taken from    Ref.~\cite{Acharya:2017fau}.}
\end{figure}

The effects of the event selection on the CME signal in this analysis were accounted for by studying the variation of $\mean{B^2\cos2(\psi_{B}-\psi_2)}$  using different initial geometry models; see Fig.~\ref{fig:ALICE_ESE_B2}. 
Here, $B$ and $\psi_{B}$ are the magnitude and azimuthal angle of the normal to the direction of the magnetic field, respectively, evaluated at the  center of the overlap region.
Within the $v_2$ ranges relevant for the ALICE analysis, the quantity $\mean{B^2\cos2(\psi_{B}-\psi_2)}$  is not constant; the variations come mainly from decorrelation of the reconstructed second-order harmonic plane $\psi_2$ relative to the magnetic field direction, not much in the magnetic field magnitude.  
To account for this effect, the relevant $v_2$ region in each centrality bin is fitted by a linear function.  
The fit function is then used to extract the CME signal in the $\dg$ measurements as functions of $v_2$ in Fig.~\ref{fig:ALICE_ESE}.  
The extracted CME signal is consistent with zero~\cite{Acharya:2017fau}, setting an upper limit of $\fcme<16$\%.
Preliminary ALICE data at 5.02~TeV set an upper limit of $\fcme<5$-6\%~\cite{Wang:2023xhn}.

\begin{figure}
  \includegraphics[width=0.8\linewidth]{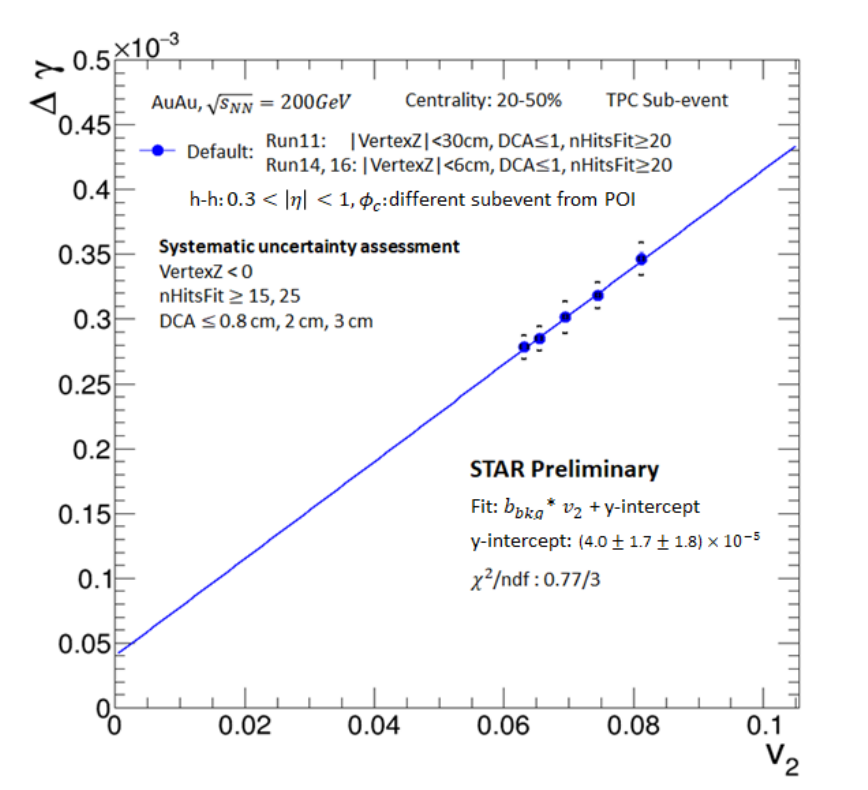} 
  \hspace*{0.4cm}\includegraphics[width=0.8\linewidth]{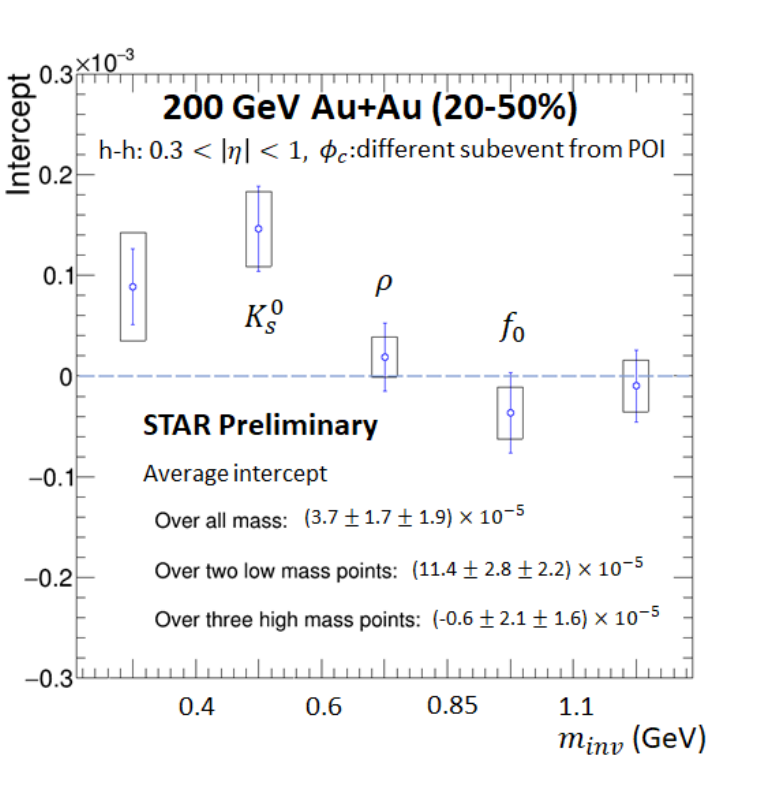}
  \caption{\label{fig:STAR_ESE}(Color online) The $\dg$ as a function
    of $v_2\two$ from an ESE analysis by STAR for the combined
    20--50\% centrality range of Au+Au collisions at $\snn=200$~GeV.
    The line is a linear fit to the data. 
    Error bars (shaded boxes) represent the statistical (systematic) uncertainties. 
    Figure is taken from Ref.~\cite{Xu:2023wcy}.}
\end{figure}

More recently, the ESE method has been applied by the STAR Collaboration~\cite{Xu:2023wcy}.  
In the STAR analysis, the TPC acceptance is divided into three subevents ($-1<\eta<-0.3$, $|\eta|<0.3$, $0.3<\eta<1$), with the middle subevent used for $q_2$ calculation. 
The $\dg$ is measured by the three-particle correlator, where the POIs are taken from  one of the two side subevents $0.3<|\eta|<1$ and particle $c$ is taken from the other side subevent.  
The $v_2\two$ is calculated by the two-particle cumulant with one particle from the subevent on one side and the other particle from the subevent on the other side.  
Figure~\ref{fig:STAR_ESE} upper panel shows the $\dg$ as a function of $v_2$ in the 20--50\% centrality range obtained from the $q_2$-selected event classes analyzed in each 10\%-size centrality bin.  
A linear fit yields an intercept corresponding to $\fcme\approx 8$\% with an $1.5\sigma$ effect.  
The analysis is repeated for several regions of restricted pair invariant mass, namely, $\minv<0.4$~GeV/$c^2$, $0.4<\minv<0.6$~GeV/$c^2$, $0.6<\minv<0.85$~GeV/$c^2$, $0.85<\minv<1.1$~GeV/$c^2$, and $\minv>1.1$~GeV/$c^2$.
The obtained intercepts are shown in the lower panel of Fig.~\ref{fig:STAR_ESE}.  
The intercept appears to be larger at low $\minv$ and consistent with zero at high $\minv$.

All ESE analyses assumed that the effects of nonflow and RP-independent background are negligible. 
These effects will have to be accounted for in future high-statistics measurements.

\subsection{Spectator/participant planes}
\label{sec:new:ppsp}

\begin{figure*}
  \includegraphics[width=0.48\linewidth]{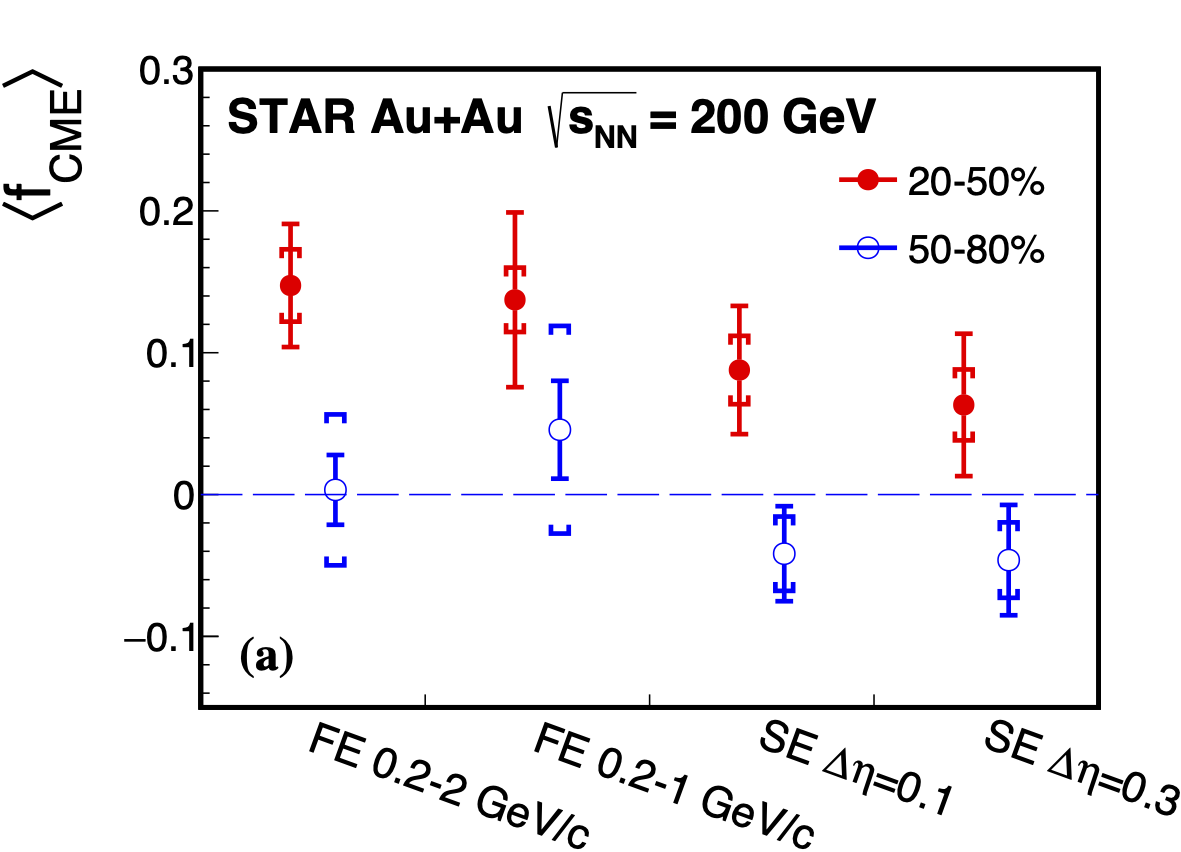}
  \includegraphics[width=0.48\linewidth]{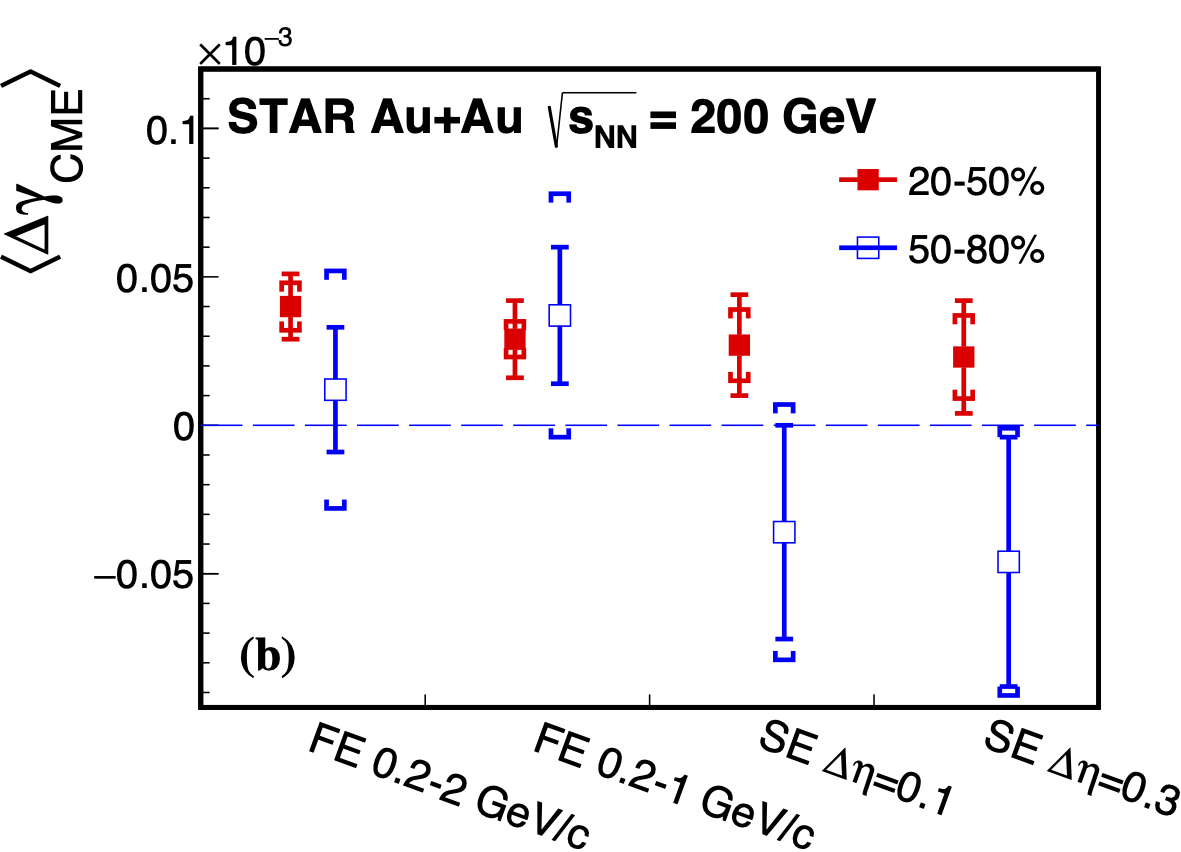}
  \caption{\label{fig:STAR_PPRP}(Color online) 
  (a) $\mean{\fcme}$ and (b) $\dg_\cme$ in 50\%–80\% (open blue markers) and 20\%–50\% (solid red markers) centrality Au+Au collisions at $\snn=200$~GeV by STAR, extracted using methods of full-event (FE) and subevent (SE) settings and kinematic cuts. 
  Error  bars and caps show statistical and systematic uncertainties, respectively. 
  Figure is taken from Ref.~\cite{STAR:2021pwb}.}
\end{figure*}

The SP/PP method, introduced in Sect.~\ref{sec:methods:ppsp}, was applied by the STAR Collaboration to Au+Au collisions at 200~GeV~\cite{STAR:2021pwb}.  
The SP was estimated by the spectator neutrons measured in the ZDC~\cite{Adams:2005ca}, and the PP by the second-order harmonic plane reconstructed from particles in the TPC.  
Figure~\ref{fig:STAR_PPRP} presents the measured CME fraction $\fcme^\obs$ in the left panel and the absolute CME signal $\dg_\cme^\obs$ in the right panel.  
Sets of four data points in each panel correspond to four different analysis settings: full-event analyses correspond to  POIs and particle $c$ taken from the full TPC acceptance $|\eta|<1$, with two POI $\pt$ ranges, and sub-event analyses correspond to selection of POIs from one side of the TPC and particle $c$ from the other side, with two sizes of the $\deta$ gap in between.  
Four results are obtained by analysis of the same data and are not statistically independent.
It is found that, while consistent with zero in peripheral 50--80\% collisions, the observed CME signal in mid-central 20--50\% collisions seems to be finite, with a 1-3$\sigma$ significance depending on the analysis cuts and methods.  
For the full-event analysis with POI kinematic range of $|\eta|<1$ and $0.2<\pt<2$~\gevc, $\fcme^\obs=(14.7\pm4.3\pm2.6)\%$, with a $2.9\sigma$ significance above zero.

This method has also been applied to Au+Au collisions at $\snn=27$~GeV.  
At that energy, the beam rapidity value is within the pseudorapidity coverage of the event-plane detector (EPD)~\cite{STAR:2022ahj}, so the inner part of EPD measures spectators and the outer part measures participant particles.  
The (inner) EPD event-plane resolution is significantly better than that of the ZDC, boosting the precision reach of the SP/PP measurement in this energy region.  
The event statistics from BES-II are still limited and the current result is consistent with zero CME signal~\cite{STAR:2022ahj}.

The SP/PP method was also used for a separate estimates of the CME signal in each of the isobar collisions~\cite{STAR:2021mii}. 
For the SP measurements, the $\gamma$ correlator and elliptic flow were measured using the event planes from the STAR's ZDCs. 
Two approaches have been used, one computes a combined first harmonic event plane from the two ZDCs and the corresponding EP resolution. 
The other uses the two $\psi_1$ harmonic planes from each of the ZDCs via
\begin{eqnarray}
(\Delta\gamma/v_2)_{\rm ZDC} 
&=& 
\frac{\Delta
  \langle\cos(\phi_{\alpha}+\phi_{\beta}-\psi_1^{\textsc{w}}-\psi_1^{\textsc{e}})\rangle}
{\langle\cos(2\phi-\psi_1^{\textsc{w}}-\psi_1^{\textsc{e}})\rangle}, 
\label{eq:dg_zdc}
\end{eqnarray}
where $\psi_1^{\textsc{w(e)}}$ is the EP determined with the ZDC on the west/backward and east/forward side of STAR, with the EP resolution extracted directly from the data as $\langle \cos(\psi_1^{\textsc{w}}-\psi_1^{\textsc{e}})\rangle$. 
While the ratio in Eq.~(\ref{eq:dg_zdc}) does not depend on the EP resolution, a quantitative estimate of $\fcme$ from the double ratio (from PP and SP) requires the $v_2$ values corrected for the EP resolutions. 
The extracted average CME fractions for the 20--50\% centrality were found to be 
consistent with zero within $\sim$15\%  uncertainties of the measurement, which are dominated by the poor ZDC EP resolution.

\subsection{Isobar collisions}
\label{sec:new:isobar}

The isobar $^{96}_{44}$Ru+$^{96}_{44}$Ru and $^{96}_{40}$Zr+$^{96}_{40}$Zr collision run was performed at RHIC in 2018.  
Tremendous efforts were put in by the BNL accelerator division to make the two isobar beam conditions as close as possible, alternating the two species on a daily basis.  
A total of approximately $2\times10^9$ good MB events (after event quality cuts) were analyzed by STAR for each isobar species, exceeding the target statistics.  
Blind analysis was performed, with the design of the data blinding scheme, the strategy of blind data calibration and analysis, and the criteria for unblinding published in Ref.~\cite{STAR:2019bjg}.  
Five independent analysis groups participated in the blind analysis with different but overlapping interests and observables.  
The most overlapped observable is the isobar ratio of the quantity $\dg/v_2$.  
The results of the analysis were published in Ref.~\cite{STAR:2021mii}.
Unprecedented precision was achieved with a statistical uncertainty of 0.4\% and a negligible systematic uncertainty on the ratio of $\dg/v_2$ between the two isobar systems.

\begin{figure*}[hbt]
  \includegraphics[width=1.02\linewidth]{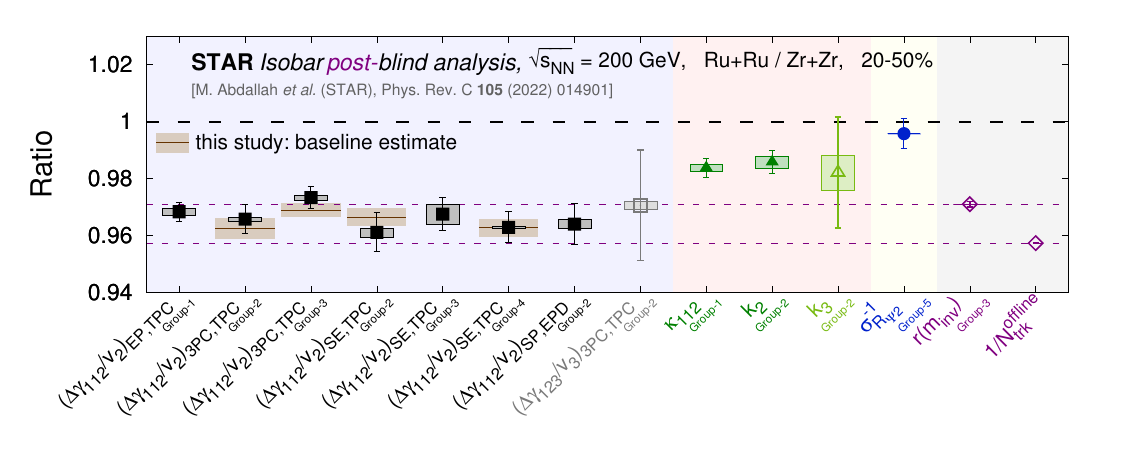}
  \caption{\label{fig:isobar}(Color online) 
  Isobar \ru/\zr\ ratios of $\dg/v_2$ from the STAR blind analyses (black squares, with error bars and gray boxes indicating statistical and systematic uncertainties, respectively)~\cite{STAR:2021mii}. Also shown are background baseline estimates by STAR for the four measurements that used the cumulant method (short horizontal lines accompanied by shaded boxes, the heights of which indicate the combined statistical and systematic  uncertainties)~\cite{STAR:2023gzg,STAR:2023ioo}. The rightmost and second rightmost purple diamonds indicate the isobar ratios of inverse multiplicities ($1/N_{\rm trk}^{\rm offline}$) and the relative pair excess of OS over SS pairs ($r$). 
  Figure is taken from Ref.~\cite{STAR:2023gzg,STAR:2023ioo}.}
\end{figure*}

Naively, for non-zero CME contribution, one would expect a larger-than-unity isobar ratio of the $\dg/v_2$ measures in \ru\ and \zr\ collisions.  
The measured ratio, see~Fig.~\ref{fig:isobar}, is smaller than unity, which is due to a few percent difference in the multiplicities between the two isobar collision systems, greater in Ru+Ru than in Zr+Zr collisions.  
The ratio of multiplicities is shown in Fig.~\ref{fig:isobar} by purple diamonds in the right part of the plot, which deviates from unity.
A difference in multiplicities was predicted by energy density functional theory calculations to root in the difference in nuclear structure between the $^{96}_{44}$Ru and $^{96}_{40}$Zr nuclei~\cite{Li:2018oec,Xu:2021vpn}.
Due to a larger number of neutrons in the $^{96}_{40}$Zr nucleus, the neutron skin is thicker.
The overall size of $^{96}_{40}$Zr is larger than that of $^{96}_{44}$Ru, resulting in a smaller energy density and hence lower multiplicity.

Because the background contribution scales with the inverse multiplicity, the baseline for the $\dg/v_2$ isobar ratio would be the inverse multiplicity ratio, as shown by the lower dashed purple line in Fig.~\ref{fig:isobar}.  
One can see that all the $\dg/v_2$ isobar ratio measurements are above this baseline, suggesting possibly a finite CME signal.  
It was shown by Ref.~\cite{Kharzeev:2022hqz} that the CME signal fraction would be $(6.8\pm2.6)\%$ if the lower dashed line was indeed the correct baseline.  
However, the number of clusters contributing to the background is not necessarily proportional to the final-state particle multiplicity with a percent accuracy.  
The relative abundances of clusters  could be estimated by $r\equiv(N_\sos-N_\sss)/N_\sss$, which is also measured by STAR as indicated by the upper dashed purple line in Fig.~\ref{fig:isobar}.
As can be seen, the $\dg/v_2$ isobar ratios are mostly below this baseline.  
Clearly, at this level of accuracy, one cannot conclude on the existence of the CME or the lack thereof. 
STAR's more accurate estimates of the background and accounting for the nonflow effects in isobar results are discussed below in Sect.~\ref{sec:new:refine}.

The initial estimates of the required statistics for the $\fcme$ measurements in isobar collisions in Ref.~\cite{Deng:2016knn} were guided by Au+Au data.
However, the relative signal strength in smaller nuclei collisions is likely to be significantly smaller. 
\comment{However, although the background scaling with inverse multiplicity is clear, the strength of chiral axial current is largely unknown.} 
In nuclear collision, the effective magnetic field scales approximately as $Z/R^2 \sim Z/A^{2/3} \sim A^{1/3}$ (where $A$ is the mass number of the colliding nuclei), so the $\dg\propto B^2$ signal would be a factor $\sim 1.5$ smaller in isobar collisions than in Au+Au collisions. 
Accounting for the larger decorrelation of the magnetic field direction with the flow planes points to even a larger decrease; see the discussion in Sect.~\ref{sec:small-large} and Fig.~\ref{fig:CME_glb} below. 
The lifetime of the magnetic field will also be shorter in isobar collisions. 
Accounting for the scaling of the background and the signal with the event multiplicity could also work in the same direction~\cite{Feng:2021oub}.
Thus, the $\fcme$ fraction in smaller nuclei collisions can be likely smaller by at least a factor of several~\cite{Feng:2021oub,ALICE:2022ljz}.
A reduction in the CME signal due to the final-state interactions, on the other hand, would likely be smaller in isobar collisions than in Au+Au collisions~\cite{Ma:2011uma,Deng:2018dut}. 
This would be a competing effect favoring isobar collisions; however, the magnitude of this effect is largely unknown.
Considering all these effects, the small $\fcme$ observed in  isobar collisions does not exclude a significant signal in collisions of larger nuclei.

\subsection{Utilizing small-large systems}
\label{sec:small-large}

\begin{figure}[tb]
    \begin{center}
    \includegraphics[width = 0.45\textwidth]{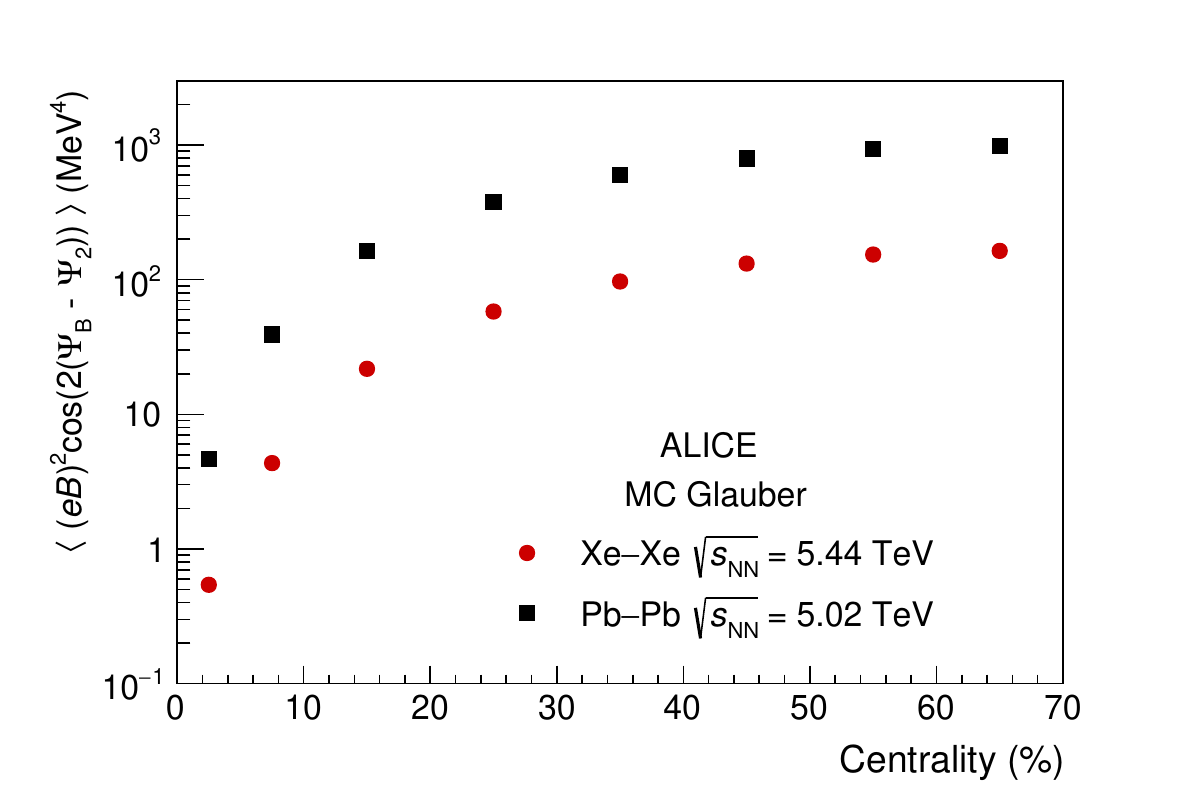}
    \end{center}
    \caption{(Color online) The expected CME signal as a function of centrality from MC Glauber simulations for Xe--Xe and Pb--Pb collisions by ALICE. Figure is taken from Ref.~\cite{ALICE:2022ljz}.} 
    \label{fig:CME_glb}
\end{figure}

The ALICE Collaboration has taken the isobar idea  further and tried to extract the CME signal from a comparison of the $\dg$ measurements in Xe+Xe and Pb+Pb collisions~\cite{ALICE:2022ljz}. 
The multiplicity scaled $\gamma$ correlators are compared since the flow-induced background is approximately inversely proportional to the multiplicity.  
Assuming that both the CME signal and the background scale with inverse multiplicity, the charge dependence of $\dg$ for the two collision systems can be expressed as
\begin{subequations}\label{eq:gg}
\begin{align}
    \Gamma^{\rm Xe-Xe} & = s \tilde{B}^{\rm Xe-Xe} + b v_{2}^{\rm Xe-Xe}, \label{eq:g1}\\
    \Gamma^{\rm Pb-Pb} & = s \tilde{B}^{\rm Pb-Pb} + b v_{2}^{\rm Pb-Pb}, \label{eq:g2}
\end{align}    
\end{subequations}
where $\Gamma \equiv \dg \, \dndeta$ (here $\dndeta$ is the pseudorapidity density of charged hadron multiplicity), and $\tilde{B} \equiv \mean{B^2 \cos2(\psi_{B}-\psi_2)}$. 
The $s$ and $b$ parameters do not depend on collision system and quantify the signal and background contributions, respectively. 
Using model calculations of $\tilde{B}$ (see Fig.~\ref{fig:CME_glb}), the $s$ and $b$ parameters can be extracted from the data using Eqs.~(\ref{eq:gg}). 
These parameters can be used to calculate the fractions of the CME signal in Xe--Xe
and Pb--Pb collisions as
\begin{equation}
    \fcme = \frac{s\tilde{B}}{s\tilde{B} + bv_{2}}.
    \label{eq:fcme2}
\end{equation}

\begin{figure}[tb]
    \begin{center}
    \includegraphics[width = 0.5\textwidth]{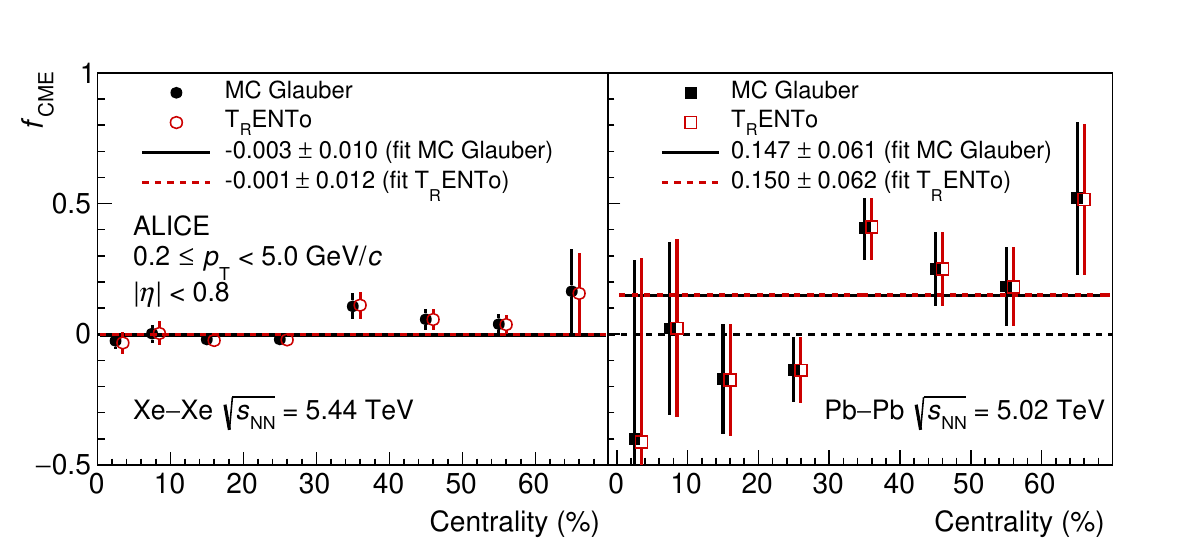}
    \end{center}
    \caption{(Color online) Centrality dependence of the CME fraction extracted by ALICE based on the expected CME signal from MC Glauber (closed markers) and T$_{\rm R}$ENTo~(open markers) models (see text for details). The T$_{\rm R}$ENTo points are slightly shifted along the horizontal axis for better visibility. Figure is taken from Ref.~\cite{ALICE:2022ljz}.}
    \label{fig:fCME_glb_tr}
\end{figure}

The results obtained with the help of Eq.~(\ref{eq:fcme2}) are shown in Fig.~\ref{fig:fCME_glb_tr}.
A smaller CME signal in Xe--Xe compared to Pb--Pb collisions is due to the smaller magnetic field strength and a larger decorrelation between $\psi_{B}$ 
and $\psi_2$. 
It is worth noting that the CME fractions in the two systems are correlated because both are calculated with the same $s$ and $b$ parameters. 
The estimated $\fcme$ is currently consistent with zero, and an upper limit has been extracted.  
Since the two systems are far more different than isobars, final-state interactions can cause a significant difference.  
ALICE has tested this allowing a large range of variations and the qualitative conclusion is unchanged.

%

\subsection{Refining results. Accounting for nonflow contribution}
\label{sec:new:refine}

The isobar analysis as well as the application of the SP/PP method to Au+Au collisions at RHIC are two the most precise measurements as of today. 
Both of them reliably isolate most of the background contribution by considering the ratios of two measurements, with one of them containing a larger CME contribution than the other. 
In one case, that is the ratio of measurements in different isobar collisions, and in another the results obtained relative to the SP and PP. 
While the dominant part of the background can be removed in this way, more subtle effects such as the difference in nonflow contributions in the two isobar collisions or nonflow effects in measurements relative to PP, and contributions from RP-independent three-particle correlations could still bias the final estimates of  $\fcme$.     
The biases due to nonflow in the $v_2$ measurements and the RP-independent three-particle correlation contribution to $\dgamma$ have very different centrality dependence. 
Nonflow contribution is minimal for midcentral collisions where the real flow is maximum.  
The RP-independent three-particle correlations are inversely proportional to the {\em square} of the multiplicity and thus are most significant in peripheral collisions. 
In principle, such a dependence could be used for testing the methods that account for the corresponding biases.

The nonflow and the RP-independent effects in SP/PP measurements have been investigated in Ref.~\cite{Feng:2021pgf}. 
Both effects affect only measurements relative to the PP, as the RP-independent correlations with the SP are expected to be negligible. 
The nonflow contribution to $\vtwopp$ increases the ratio $\dgsp/\dgpp$, resulting in a positive contribution to the $\fcme$ estimate.
The three-particle RP-independent contamination gives a negative contribution to the $\fcme$ estimate because it increases $\dg\tpc$ making its difference from $\dg\zdc$ smaller. 
Thus, the two effects partially cancel each other.

\begin{figure*}
  \includegraphics[width=0.47\linewidth]{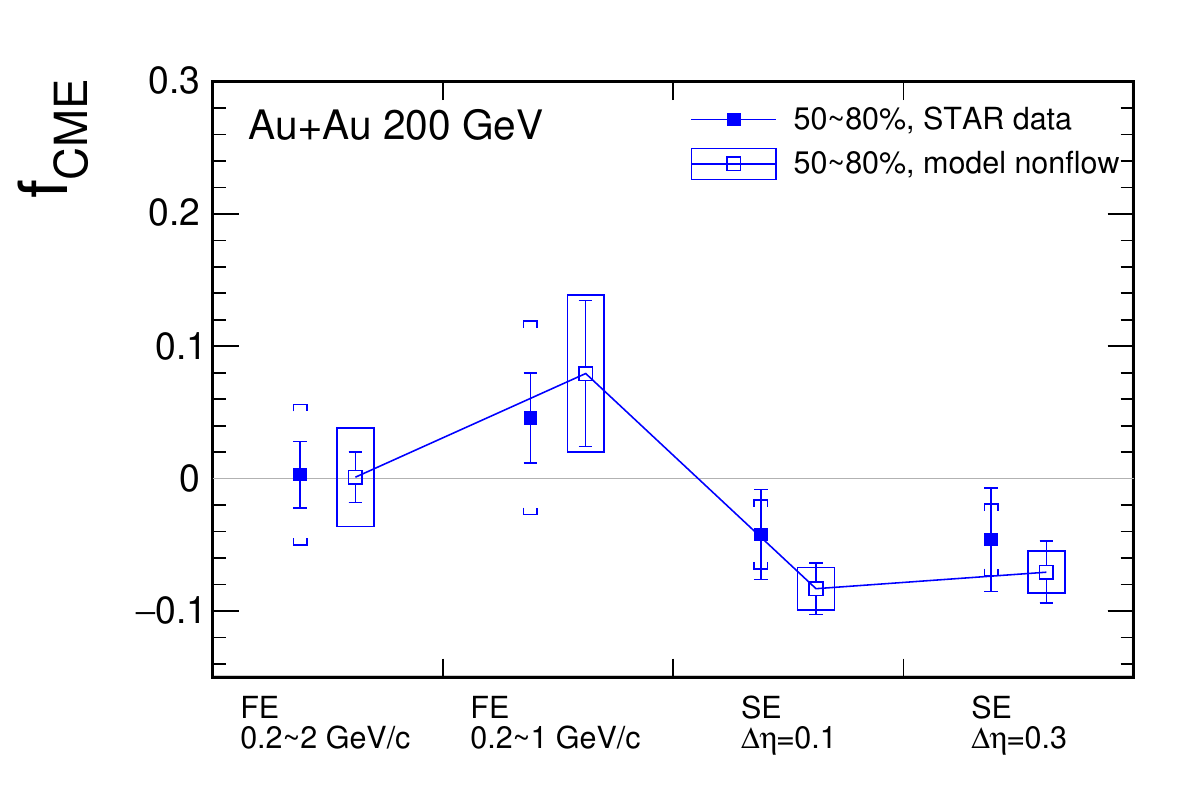}
  \includegraphics[width=0.47\linewidth]{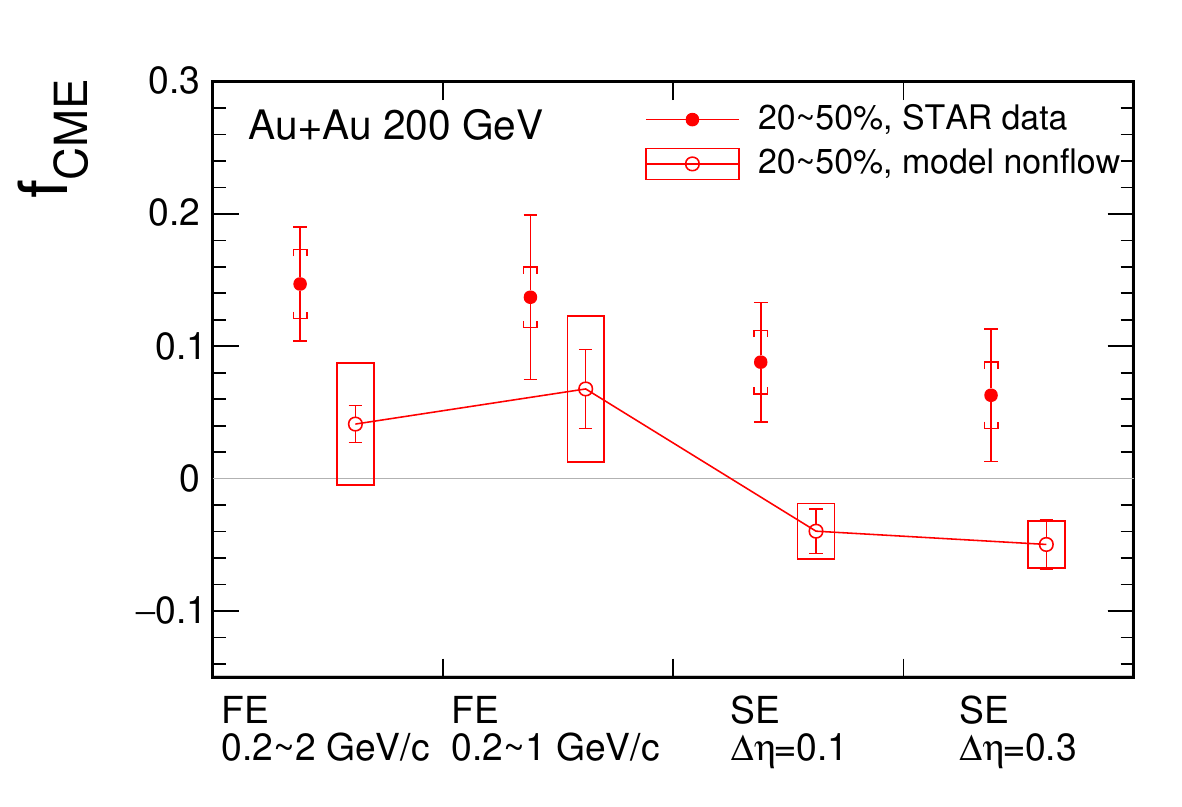}
  \caption{\label{fig:STAR_fcme}(Color online) The CME fraction $\mean{\fcme}$ measured by STAR~\cite{STAR:2021pwb}
  compared to model estimates of nonflow  contributions by  Ref.~\cite{Feng:2021pgf} in peripheral 50--80\% (left panel) and central 20--50\% (right panel) Au+Au collisions at    $\snn=200$~GeV. 
  Error bars show statistical  uncertainties; the caps (boxes) indicate the systematic
    uncertainties of STAR data (model estimates of nonflow contributions). 
    Figure is reproduced from Ref.~\cite{Feng:2021pgf}.}
\end{figure*}

In Ref.~\cite{Feng:2021pgf} the nonflow effect in $v_2$ was estimated by using the \ampt\ model.  
The effect of RP-independent three-particle correlations is estimated by using the \hijing\ model. 
The estimated nonflow contribution to the observed CME fraction, $\fcme^{\rm model}$, is shown in Fig.~\ref{fig:STAR_fcme} by the points connected by lines, compared to the STAR measurement of $\fcme^\obs$.  
It is found that the net effect of nonflow correlations causes a nearly zero or even negative bias to $\fcme$, dependent of the analysis cuts and methods employed in the STAR data analysis~\cite{STAR:2021pwb}.  
The difference $\fcme=\fcme^\obs-\fcme^{\rm model}$ may be regarded as a refined estimate of the CME signal, which suggests possibly a finite CME signal in mid-central collisions. 
It will be interesting to have a more data-driven analysis of background effects in the future.

Experimentally, the measurement of elliptic flow relative to the SP, usually done with ZDCs,  might be rather challenging, as it involves a good knowledge of the ZDC EP resolution. On the other hand, the ratio  $\frac{\dgv^\ssp}{\dgv^\spp}$ does not require the knowledge of the EP resolutions (see Eq.~\ref{eq:Group4_fcmeSP}). 
This ratio might be used first for a  statistically more accurate estimate of the presence of the CME, while its magnitude determination would still require $\vtwosp/\vtwopp$, which in turn requires the knowledge of EP resolutions.

An attempt to account for nonflow effects in an estimate of the ``baseline'' in the isobar comparison has been made by the STAR Collaboration in Refs.~\cite{STAR:2023gzg,STAR:2023ioo}. 
A 2D fitting method of two-particle correlations in $(\deta,\dphi)$ was used to estimate the (difference) in nonflow contamination in $v_2\two$ measurements.  
The systematic uncertainties are assessed by allowing longitudinal flow decorrelation~\cite{Bozek:2010vz,Xiao:2012uw,CMS:2015xmx,ATLAS:2017rij} and varying the functional form for nonflow correlations.
The three-particle nonflow contamination was estimated by the \hijing\ model. 
\hijing\ does not include collective flow, so its entire three-particle correlations can be used as an estimate of the RP-independent background~\cite{Wang:1991xy,Wang:1998bha}.  
The systematic uncertainty is assessed by \hijing\ with jet-quenching turned off.
It was found that nonflow contribution to $v_2$ is smaller in Ru+Ru.  
Similarly, the three-particle correlation background was also found smaller in Ru+Ru collisions. 
Both effects are presumably due to the larger multiplicity dilution in Ru+Ru collisions. 
The two effects partially cancel each other, and the net effect turns out to be slightly negative.
The estimated background baselines for four of the $\dg/v_2$ isobar ratio measurements, where the cumulant method is used in measuring $v_2\two$, are shown by brown lines accompanied by shaded bars in Fig.~\ref{fig:isobar}.  
The measurements of the $\dg/v_2$ isobar ratios are consistent with the estimated background baselines, within approximately $1\sigma$ uncertainty.  
An upper limit of approximately 10\% for $\fcme$ is extracted for each of the four measurements at the confidence level of 95\%.  
For reference, the highest precision data point of the $\dg/v_2$ isobar ratios in Fig.~\ref{fig:isobar} has a statistical uncertainty of 0.4\%, which corresponds to an uncertainty of approximately $3\%$ in terms of $\fcme$.

\section{Summary and Outlook}

The chiral magnetic effect (CME) is a phenomenon of fundamental importance.  
It manifests itself as an electric current along a magnetic field in  domains of imbalanced left- and right-handed quarks and antiquarks.  
Such domains arise from vacuum fluctuations, that  violate the parity and charge-parity symmetries, and root in 
the basic properties of quantum chromodynamics, the fundamental theory of the strong interaction.  
Relativistic heavy-ion collisions appear to be a perfect place to search for the CME, as they possess the strongest magnetic field possible in an experiment and the vacuum fluctuations could be enhanced due to high temperature achieved in those collisions. 
The experimental searches for the CME at the Relativistic Heavy-Ion Collider (RHIC) and the Large Hadron Collider (LHC) have  so far been inconclusive due to large background effects.  
The major background comes from charge-dependent particle correlations modulated by the elliptic flow, causing a positive contribution to the correlator $\dg$~\cite{Voloshin:2004vk}, an observable widely used in CME searches.  
Subleading backgrounds arise from correlations not related to the reaction plane; these  have to be accounted for in order to achieve the accuracy in $\fcme$ at the percentage level. 

The RHIC isobar program of $^{96}_{44}$Ru+$^{96}_{44}$Ru and $^{96}_{40}$Zr+$^{96}_{40}$Zr collisions, designed to address the background issue, did not lead to a firm conclusion about the CME, despite its successful execution beyond the expected quality and statistics~\cite{STAR:2021mii}.  
The difficulty arises from the reduced signal-to-background ratio in relatively small systems compared to the heavier Au+Au or Pb+Pb collisions and subtle differences in the nuclear structure of the $^{96}_{44}$Ru and $^{96}_{40}$Zr nuclei, that become important for small signals.
An upper limit of approximately 10\% CME signal in the measured $\dg$ in isobar collisions is extracted at 95\% confidence level~\cite{STAR:2023gzg,STAR:2023ioo}.  
It should be pointed out that the isobar result is not a disproof of the CME but provides  guideline for its possible magnitude.  
Collisions of higher-mass isobars would be a viable venue for future searches of the CME.

On the other hand, an intriguing hint of a possible CME signal of the order of 15\% of the measured $\dg$ has been found with a significance of $2.9\sigma$ in midcentral Au+Au collisions at $\snn=200$~GeV at RHIC~\cite{STAR:2021pwb}.  
This is in line with the expectation that the CME signal-to-background ratio is at least a factor of 3 larger in Au+Au than in isobar collisions. 
The searches for the CME at the LHC have so far only allowed the extraction of upper limits, where the CME signal may be weaker because of the faster decrease of the magnetic field at the higher energies.

In CME searches, it is crucial to pin down the backgrounds, because of their major contributions, in model-independent data-driven ways. 
Background isolation should be unambiguous with clearly interpretable systematics.  
In our view, barring a future high-mass isobar program, there are two known methods that could achieve this goal: the event-shape engineering with an event-shape variable and particles of interest well distanced in momentum space, and the spectator/participant planes method.  
The former has been analyzed by LHC experiments with wide acceptance detectors and should become viable with new and future RHIC data with increased forward detector capabilities.  
The latter method has been used at RHIC owing to the good granularity of the shower maximum detectors of the zero-degree calorimeters (ZDC) and might see future applications at the LHC with improved ZDC detectors.  

The CME search has so far concentrated on using unidentified charged hadrons and the primary $\dg$ observable. It would be valuable also to use identified hadrons which may shed more light on the physics of different contributions to $\dg$. 
Correlating other CME-sensitive measures to $\dg$~\cite{Du:2008zzb,Finch:2018ner,STAR:2023qyt} can also be powerful. Future studies along these lines would be interesting and important.

It is anticipated that a factor of 10 or more increase in event statistics will be achieved in Au+Au collisions at RHIC by the end of 2025.
The upgraded detector of larger rapidity coverage and the newly installed forward event-plane detector in the STAR experiment will boost the CME search capability beyond the mere increase in the event statistics.  
Considering the intriguing hint of a possible CME signal in the current Au+Au data, the expected 10-fold increase in event statistics, and the enhanced detector capabilities, we remain optimistic about the CME measurements in the near future.
